\documentclass[a4paper]{article}
\pdfoutput=1
\usepackage{jheppub}

\usepackage{amsmath}

\usepackage{amssymb}
\usepackage{graphicx,color}

\usepackage{amsmath,amsthm}
\usepackage{txfonts}

\usepackage{mathrsfs}

\usepackage{bm}

\newcommand{\Ref}[1]{(\ref{#1})}

\newcommand{\R}{\mathbb{R}}

\newcommand{\half}{\frac{1}{2}}

\newcommand{\bA}{{\bar{A}}}

\newcommand{\Su}{\mathrm{SU}(2)}

\def\be{\begin{eqnarray}}
\def\ee{\end{eqnarray}}

\newcommand{\ca}{\mathcal A}

\newcommand{\cc}{\mathcal C}

\newcommand{\ce}{\mathcal E}
\newcommand{\cf}{\mathcal F}
\newcommand{\cg}{\mathcal G}
\newcommand{\ch}{\mathcal H}
\newcommand{\ci}{\mathcal I}

\newcommand{\cn}{\mathcal N}

\newcommand{\cp}{\mathcal P}

\newcommand{\calr}{\mathcal R}
\newcommand{\cs}{\mathcal S}

\newcommand{\cv}{\mathcal V}

\newcommand{\fp}{\mathfrak{p}}

\renewcommand{\b}{\beta}
\newcommand{\g}{\gamma}
\newcommand{\G}{\Gamma}

\newcommand{\Sig}{\Sigma}

\renewcommand{\L }{\Lambda}
\renewcommand{\o}{\omega}

\newcommand{\rmd}{\mathrm d}

\newcommand{\lt}{\left}
\newcommand{\rt}{\right}

\newcommand{\lag}{\left\langle}
\newcommand{\rag}{\right\rangle}

\newcommand{\tr}{\mathrm{tr}}

\newcommand{\Ar}{\mathbf{Ar}}

\title{Loop Quantum Gravity, Exact Holographic Mapping, and Holographic Entanglement Entropy}

\author[1,2]{Muxin Han}

\author[3,4,5]{,\ \ Ling-Yan Hung}

\affiliation[1]{Department of Physics, Florida Atlantic University, 777 Glades Road, Boca Raton, FL 33431-0991, USA}

\affiliation[2]{Institut f\"ur Quantengravitation, Universit\"at Erlangen-N\"urnberg, Staudtstr. 7/B2, 91058 Erlangen, Germany}

\affiliation[3]{Department of Physics and Center for Field Theory and Particle Physics, Fudan University, 220 Handan Road, 200433 Shanghai, China}

\affiliation[4]{State Key Laboratory of Surface Physics and Department of Physics, Fudan University, 220 Handan Road, 200433 Shanghai, China}

\affiliation[5]{Collaborative Innovation Center of Advanced Microstructures, Nanjing University, Nanjing, 210093, China}

\abstract{The relation between Loop Quantum Gravity (LQG) and tensor network is explored from the perspectives of bulk-boundary duality and holographic entanglement entropy. We find that the LQG spin-network states in a space $\Sig$ with boundary $\partial\Sig$ is an exact holographic mapping similar to the proposal in \cite{Qi2}. The tensor network, understood as the boundary quantum state, is the output of the exact holographic mapping emerging from a coarse graining procedure of spin-networks. Furthermore, when a region $A$ and its complement $\bA$ are specified on the boundary $\partial\Sig$, we show that the boundary entanglement entropy $S(A)$ of the emergent tensor network satisfies the Ryu-Takayanagi formula in the semiclassical regime, i.e. $S(A)$ is proportional to the minimal area of the bulk surface attached to the boundary of $A$ in $\partial\Sig$. 

}


\arxivnumber{}

\begin{document}

\maketitle

\section{Holography, Tensor Network, and Loop Quantum Gravity}

The AdS/CFT correspondence has brought about new surprises and insights in deciphering the nature of gravity and particularly quantum gravity. The new perspective that brought about these advances was the study of entanglement entropy in field theories, and the understanding of its manifestation in the gravity dual. It was first conjectured by Ryu and Takayanagi  (RT) \cite{Ryu:2006bv} that the entanglement entropy of some chosen region A in configuration space in the CFT is proportional to the area of some minimal surface in the AdS dual homologous to the boundary of this region at the AdS boundary. Explicitly,
\be
S_{EE}(A) = \frac{\Ar_{\textrm{min}}}{4G_N},
\ee
where $G_N$ is the Newton's constant. This formula has since been checked in many non-trivial examples, and subsequently proved in 1+1 dimensions\cite{Faulkner:2013yia} and then higher dimensions \cite{Lewkowycz:2013nqa} using other established techniques in the  AdS/CFT correspondence. Long before these works however, it is suspected \cite{Ryu:2006bv} and later confirmed\cite{Casini:2011kv} that the formula for holographic entanglement entropy is closely related to black hole entropy, coinciding with the Bekenstein Hawking formula in some simple cases, when the region $A$ is one bounded by a sphere $S^{d-1}$. As it is well known, there are many works in the past attempting to explain the  black hole entropy as a gravitational entanglement entropy \cite{Sorkin:2014kta,Callan:1994py}. Recent works have also demonstrated that this entanglement entropy of the boundary is itself closely related to the entanglement entropy of the bulk gravity theory\cite{Faulkner:2013ana,Jafferis:2015del, Almheiri:2014lwa,Harlow:2016vwg}. One rather universal feature in all these discussions is the emergence of area law being a, if not \emph{the}, crucial signature of a local semi-classical gravitational background\cite{Bianchi:2012ev}. 

\begin{figure}
\begin{center}
\includegraphics[width = 0.5\textwidth]{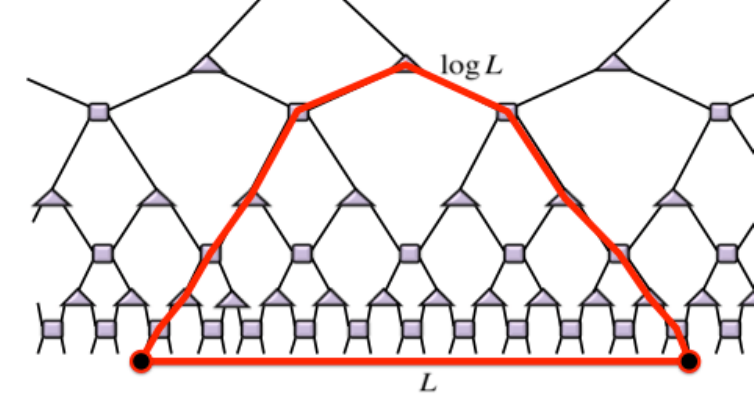}
\end{center}
\caption{Entanglement entropy of a region is bounded by a minimal curve (red curve) cutting through the MERA tensor network. (Figure adapted and modified from  \cite{Orus:2014poa})}
\label{fig:Mera}
\end{figure}

One very natural question that arises is : where does the RT formula come from? What is the underlying structure that can give rise to an area law? Swingle made a profound observation in \cite{Swingle:2009bg}. Working on tensor networks in condensed matter aimed at obtaining in a numerically efficient way the ground states of interacting many body Hamiltonian, it is observed that the entanglement of ground state wavefunctions describable by MERA type tensor networks carries an entanglement entropy that is proportional to the number of links on a minimal cut through the tensor network. The picture is highly suggestive of the RT formula, as shown in Figure \ref{fig:Mera}. 

\begin{figure}
\begin{center}
\includegraphics[width = 0.48\textwidth]{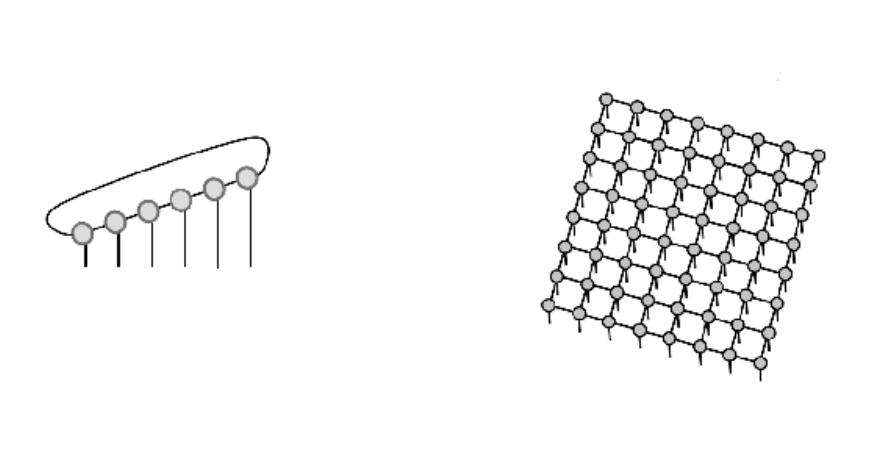}
\end{center}
\caption{Various examples of tensor networks: MPS on the left and  PEP on the right. (Figure adapted and modified from \cite{2013PhRvB..88k5147S})}
\label{fig:tensors}
\end{figure}

The observation has inspired a lot of effort aiming to explore how the gravitational bulk/tensor network correspondence should play out. More precisely, tensor networks are general ways of re-writing a many-body wavefunction in terms of contraction of tensors.  Explicitly, 

\be
|\Psi\rangle = \sum_{\{a_i\}}f_{a_1,a_2,\cdots a_N} |a_1,a_2,\cdots, a_N\rangle = \sum_{\{a_i, \gamma_l\}}\prod_I T^I_{\gamma_1\, \gamma_2 \,...\, a_i...}   |a_1,a_2,\cdots a_N\rangle, 
\ee
where $f_{a_1,a_2,\cdots a_N}$ is the amplitude of a particular state, which has been rewritten in terms of the contraction of tensors $T^I_{\gamma_1\, \gamma_2 \,...\, a_i...} $, with the $\gamma_i$ indices auxiliary ``internal indices'' that are contracted among these $T^I$'s forming a network. The indices $a_i$, the physical indices, remain uncontracted among tensors, and are often  called the ``dangling legs''. The superscript $I$ is a label for these tensors, denoting different choices of tensors that can be placed in the network so that the energy wrt a given Hamiltonian is minimized. Different architecture of these tensors serve different purposes : the MPS tensor network for example, are very efficient modeling ground states of gapped Hamiltonians, whereas MERA is specialized in capturing ground states of gapless systems\cite{Bridgeman:2016dhh}. The key lies in the fact that the architecture of the tensor network can be viewed as a process of real-space renormalization, so that global symmetries of the wavefunction -- scaling symmetry for example-- is encapsulated in the geometry of the tensor network \cite{2008PhRvL.101k0501V}.  Various example of the tensor network other than MERA are shown in Figure \ref{fig:tensors}. As observed in Swingle\cite{Swingle:2009bg}, the entanglement is indeed generically bounded by the number of legs on the analogue of an RT surface homologous to the boundary ``dangling legs'' cutting through the network, which we have mentioned above. 

An upper bound is a relatively vague statement. To make further progress, it is observed in \cite{Almheiri:2014lwa} and subsequently constructed explicitly in \cite{Pastawski:2015qua} that the RT formula, together with the causal structure in classical AdS space implies that a tensor network that re-enacts the gravity theory is probably one that behaves like an error correcting code\footnote{An error correcting code refers to protocals adopted to encode information such that at least some specific forms of error in the process of the encoding can be detected and subsequently corrected. This is achieved by introducing more bits than the actual number of bits of information that is encoded. In the case of a quantum error correcting code, we introduce extra qubits to encode each bit of the actual information, which is called logical bit. The logical bit would thus occupy only a subspace of the full Physical Hilbert space, which is called a logical subspace, and error detection is achieved by determining whether a state has been taken outside of the logical subspace. The AdS/CFT correspondence resembles some features of the error correcting code in the sense that information of the semi-classical bulk can be recovered requiring only some parts of the boundary and not all. Therefore bulk information appears like logical qubits, whereas the CFT states behave like physical bits. } built from perfect tensors, which has the property that it is a unitary map from any choice of half of its full set of indices to the other half. Further works\cite{Qi1} uncover the natural place in which perfect tensors can be found -- generic tensors with large bond dimensions. 

At this point, a very interesting picture emerges. Now that tensor networks prove itself a promising candidate for describing crucial features of (quantum) gravity, the physical picture that it presents is highly reminiscent of the picture that is intrinsic to the loop quantum gravity program. In this paper, we would like to make the connection between tensor networks and the loop quantum gravity framework explicit. Or in other words, it is possible to frame the loop quantum gravity program using perspectives of the tensor network construction, so that beautiful features such as the RT formula resurfaces naturally in the loop quantum gravity program.

Loop quantum gravity (LQG) is an attempt toward a nonperturbative and background independent quantum theory of gravity in 3, 4, and higher spacetime dimensions \cite{book,rovelli2014covariant,review1,review,Bodendorfer:2011nx}. In this paper we mainly focus on 4 spacetime dimensions. LQG is originated by the canonical formulation of classical gravity in 4d as a dynamical theory of gauge connections \cite{Ashtekar:1986yd}. In this formalism, the phase space $\cp$ of gravity has a similar structure as an SU(2) gauge theory. But the canonical variables represent the geometry on 3d spatial slices. The quantization of the phase space $\cp$ has been well-understood in LQG since 1990s, see e.g. \cite{Rovelli1988,Ashtekar:1991kc,book}. It leads to the Hilbert space $\ch_{LQG}$, shown to be the \emph{unique} representation of the operator algebra quantizing the phase space $\cp$ \cite{uniqueness,Fleischhack:2007mj}. Promoting the canonical variables to operators on $\ch_{LQG}$ quantizes the 3d spatial geometry. Many geometrical quantities are represented as (self-adjoint) operators on $\ch_{LQG}$ (e.g. \cite{Rovelli1995,ALarea,ALvolume,Thiemann:1996at,Bianchi:2008es,Ma:2010fy,Ma:2000au}). Two of the most important examples are the area operator $\hat{\Ar}_S$ and the volume operator $\hat{V}_\calr$. Both of them have the discrete spectra (eigenvalues), which implies that in quantum geometry, the area and volume are fundamentally discrete at Planck scale. The area and volume operators share the same set of eigenstates in $\ch_{LQG}$, which are known as \emph{spin-network} states. The area and volume eigenvalues are understood as the quanta carried by the spin-networks at Planck scale.

The purpose of this paper is to explore the relation between LQG and tensor network, especially in the perspectives of bulk-boundary duality and holographic entanglement entropy. We find that the LQG spin-network states on a space $\Sig$ with boundary $\partial\Sig$ is an exact holographic mapping similar as the proposal in \cite{Qi2,Qi1}. The tensor network, understood as the boundary quantum state, is the output of the exact holographic mapping emerging from a coarse graining procedure of spin-networks. Furthermore, when a region $A$ and its complement $\bA$ are specified on the boundary $\partial\Sig$, we show that the entanglement entropy $S(A)$ of the resulting tensor network satisfies the Ryu-Takayanagi formula in the semiclassical regime, i.e. $S(A)$ is proportional to the minimal area of the bulk surface attached to the boundary of $A$ in $\partial\Sig$.

\begin{figure}[h]
\begin{center}
\includegraphics[width=15cm]{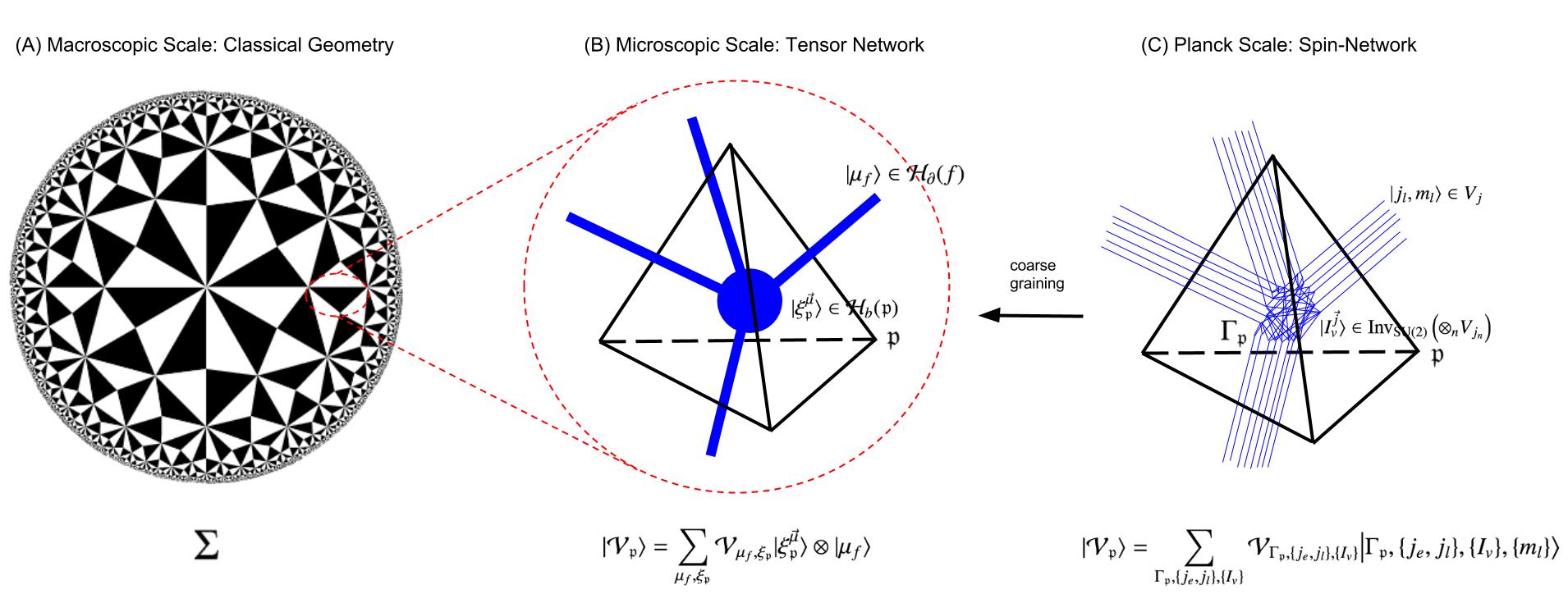}
\caption{The spatial region $\Sig$ with boundary $\partial\Sig$ and its semi-classical geometry are built by a large number of polyhedra $\fp$ (tetrahedra shown in figure) with semi-classical geometry. The semi-classical geometry of $\fp$ is fundamentally made by the spin-networks $|\G_\fp,\{j_e,j_l\},\{I_v\},\{m_l\}\big\rangle$ with a large number of edges and vertices in the graph $\G_\fp$, shown in Figure (C). Each edge carries a spin $j$ as the quanta of area at Planck scale, while each vertex carries a intertwiner $I_v$ as the quanta of volume. The spin-networks with the large number of degree of freedom can be coarse grained to the picture shown in Figure (B). Each of the 4 legs in Figure (B) represents the Hilbert space $\ch_\partial(f)$, whose basis is labelled by $|\mu_f\rangle$. $\ch_\partial(f)$ includes all microstates $|j_l,m_l\rangle$ (the boundary microstates of $\fp$) carried by the dangling edges in spin-network graph $\G_\fp$. The middle ball in Figure (B) represents the Hilbert space $\ch_b(\fp)$, whose basis is labelled by $|\xi^{\vec{\mu}}_\fp\rangle$. $\ch_b(\fp)$ includes all the spins on the internal edges of $\G_\fp$ and intertwiners on all vertices (the bulk microstate inside $\fp$). The coarse grained spin-networks give the exact holographic mapping, and leads to the tensor network representing the ground state of boundary CFT. Figure (A), (B), and (C) figures are the physical pictures at 3 different scales: (A) is at the macroscopic scale, where the typical length scale $L$ is the mean curvature radius of the semi-classical geometry. (B) is at the microscopic scale, where the tensor network lives, and the typical (squared) length scale is the mean face area $\Ar_f$ of polyhedron. (C) is at Planck scale, where spin-network lives, and the typical length scale is the Planck length $\ell_P$. The semiclassical regime of LQG is given by $\ell_P^2\ll A_f\ll L^2$. In this regime, we reproduce correctly the Ryu-Takayanagi formula of holographic entanglement entropy. }
\label{micro}
\end{center}
\end{figure}

One of the aims in this work is to construct LQG states, which realizes the bulk-boundary duality and relates to the tensor networks. The idea of construction is illustrated in FIG.\ref{micro}. Given the spatial region $\Sig$ discretized into a large number of polyhedra $\fp$, we introduce 3 different length scales: 

\begin{description}

\item (A) Macroscopic scale: It is the scale at which the smooth classical geometry is seen on $\Sig$. The scale is characterized by $L$ being the mean curvature radius of the geometry.  

\item (B) Microscopic scale: It is the scale of each polyhedron $\fp$, it is also the scale at which we define the quantum states as exact holographic mapping and tensor networks. The scale is characterized by the (square-root) of the mean face area $\Ar_f$ of polyhedra $\fp$.

\item (C) Planck scale $\ell_P$: It is the scale at which the spin-network states are defined. By LQG, A spin-network state associates with a network graph $\G$ consisting of a number of edges $e$ and vertices $v$. Each edge carries a Planck scale area, while each vertex carries a Planck scale volume. The areas and volumes carried by the spin-network are the eigenvalues of the area and volume operators.  

\end{description} 
The analysis in this paper focuses on the regime that  
\be
\ell_P^2\ll \Ar_f\ll L^2.\label{lAL}
\ee
This regime has been studied as the semiclassical regime of LQG from several different perspectives \cite{Ghosh:2013iwa,hanBH,Han:2016fgh,lowE,Sahlmann:2001nv}. Here $\Ar_f\ll L^2$ means that the smooth classical geometry is a good approximation when we zoom out to the macroscopic level. The discreteness from the small polyhedra $\fp$ is negligible. The tensor network $\Psi$ is build according to the polyhedral discretization of $\Sig$ at the scale $\Ar_f$. Namely each polyhedron $\fp$ associates to a tensor $T(\fp)$ whose indices associate to the polyhedron faces. Gluing a pair of polyhedra $\fp,\fp'$ corresponds to contracting a pair of indices from  $T(\fp)$ and $T(\fp')$. 

Each tensor $T(\fp)$ is understood as a quantum state in a Hilbert space $\otimes_f\ch_{\partial}(f)$, where each $\ch_{\partial}(f)$ associates to a face $f\subset\partial\fp$. We may enlarge this Hilbert space and include some bulk degrees of freedom, i.e. a bulk Hilbert space $\ch_b(\fp)$ is defined and tensor product with $\otimes_f\ch_{\partial}(f)$. A state $|\cv_\fp\rangle$ is picked up in the enlarged Hilbert space, and gives $T(\fp)$ by taking (partial) inner product with certain bulk state $|\phi_b(\fp)\rangle\in\ch_b(\fp)$, i.e. $\langle\phi_b(\fp)|\cv_\fp\rangle=|T(\fp)\rangle$. One may view $|\cv_\fp\rangle$ as an enlarged tensor $(\cv_\fp)_{\mu_b;\{\mu_f\}}$ with both the boundary index $\mu_f$ of $\ch_{\partial}(f)$ and the bulk index $\mu_b$ of $\ch_b(\fp)$. Its inner product with $|\phi_b(\fp)\rangle$, which has only bulk index $\mu_b$, gives the tensor $T(\fp)_{\{\mu_f\}}$ with only boundary indices. The tensor network $\Psi$ is made by contracting the $\mu_f$ indices of $T(\fp)$'s associated to internal $f$'s. $\Psi$ is understood as a boundary quantum state associated to $\partial\Sig$, because the un-contracted $\mu_f$'s of $\Psi$ only associate to $\partial\Sig$. But we can again define the state enlarged from $\Psi$ by adding bulk degrees of freedom. It can be obtained by define a tensor network from $|\cv_\fp\rangle$ by only contracting the $\mu_f$ indices associated to the internal $f$, leaving the bulk $\mu_b$'s un-contracted. The resulting tensor network is denoted by $|\Sig\rangle$ which has the un-contracted indices associated to both $\partial \Sig$ and the bulk of $\Sig$. The original tensor network $\Psi$ can be obtained by the partial inner product with certain bulk states $\Phi_b$ which has only bulk $\mu_b$ indices 
\be
\langle\Phi_b|\Sig\rangle=|\Psi\rangle.\label{EHM0}
\ee
The enlarged tensor network state $|\Sig\rangle$ is referred to as the \emph{exact holographic mapping}, proposed in \cite{Qi2,Qi1}. It is proposed as a concrete realization of the bulk-boundary duality. Given the exact holographic mapping $|\Sig\rangle$, the boundary quantum state is uniquely determined by the bulk quantum state.  

Coming back to the regime Eq.\Ref{lAL}, the Planck scale $\ell_P$ where LQG spin-networks live is much smaller than the scale $\Ar_f$ where the tensor-networks $|\Psi\rangle$ or exact holographic mappings $|\Sig\rangle$ live. It suggests that the spin-network states should play a more fundamental role. The tensor network and exact holographic mapping is emergent from the spin-networks. Indeed we propose that the tensor network and exact holographic mapping are obtained by spin-network states via a coarse graining procedure. 

In the exact holographic mapping $|\Sig\rangle$, a polyhedron $\fp$ associates to a state $|\cv_\fp\rangle$. But we understand $|\cv_\fp\rangle$ as a coarse grained prescription of some complicated spin-network states in $\fp$. The spin-networks in $\fp$ generically have a large number of edges and vertices. There are a large number of internal edges $e$ inside $\fp$, and a large number of dangling edges $l$ intersecting the faces of $\fp$. A large number of micro-degree of freedom are carried by the spin-network edges and vertices. 

A spin-network state $|\G,\{j_e,j_l\},\{I_v\},\{m_l\}\rangle$ is labelled by (1) a graph $\G$ consisting a number of edges $e$ and vertices $v$, (2) an SU(2) irrep $j_e\in \mathbb{N}/2$ carried by each internal edges $e$, (3) an SU(2) state $|j_l,m_l\rangle$ in irrep $V_{j_l}$ carried by each dangling edges $l$, and (4) an SU(2) invariant tensor (intertwiner) $I_v\in\mathrm{Inv}(\otimes_i V_{j_i})$ at each vertex $v$ with $j_i$'s carried by the adjacent edges. The quantum area carried by $e$ relates to $j_e$ by $\Ar_e=8\pi\g\ell_P^2\sqrt{j_e(j_e+1)}$ (the same for $l$). $\Ar_e$ is understood as the (Planck scale) area element of the surface transverse to $e$. The quantum volume $V_v$ carried by $v$ relates to both $I_v$ and the adjacent $j$'s. $V_v$ is understood as the volume element of the neighborhood at $v$. The expression of $V_v$ can be found in e.g. \cite{Brunnemann:2007ca,Bianchi:2011ub}

The spin-network states in $\fp$ describe the quantum fluctuation of polyhedral geometry at the deep Planck scale. The spin-networks and their linear combinations have both boundary and bulk micro-degrees of freedom. The boundary microstates are the states $|j_l,m_l\rangle$ at all dangling edges. Each dangling edge $l$ intersects a face of $\fp$. The bulk microstates are the internal $j_e$ and $I_v$. At the coarse grained level, the boundary and bulk microstates are grouped into the Hilbert spaces $\ch_\partial(f)$ and $\ch_b(\fp)$. The tensor index $\mu_f$ of $|\cv_\fp\rangle$ counts the microstates at all $l$'s which intersects $f$, while $\mu_b$ counts all the bulk microstates.

The exact holographic mapping $|\Sig\rangle$ is then constructed by contracting the $\mu_f$ indices of $|\cv_\fp\rangle$'s when gluing tetrahedra $\fp$'s. It effectively connects the spin-network states from each $\fp$, and consistently produces the spin-network states in the entire space $\Sig$. The resulting exact holographic mapping $|\Sig\rangle$ belongs to the LQG Hilbert space $\ch_{LQG}$ on $\Sig$, as a linear combination of spin-network states. Here importantly, $|\Sig\rangle$ generically sums all possible quantum geometries on $\Sig$.

Since $\Sig$ has the boundary $\partial\Sig$, the state $|\Sig\rangle$ has both the bulk and boundary degrees of freedom. Naively $\ch_{LQG}$ where $|\Sig\rangle$ lives may be viewed as a tensor product Hilbert space $\ch_b\otimes\ch_\partial$ of bulk and boundary microstates. However as an important feature implied by LQG, $|\Sig\rangle$ exhibits certain entanglement between bulk and boundary microstates. The reason is simple: the SU(2) intertwiner $I_v$ as bulk microstate depends on the adjacent spins $j$, which involves the boundary microstates when some adjacent edges of $v$ connect to $\partial\Sig$. Therefore the Hilbert space $\ch_{LQG}$ doesn't factorize as a tensor product of bulk and boundary. By the same reason, for each $\fp$, the state $|\cv_\fp\rangle$ is also an entangled state, whose Hilbert space doesn't factorize into bulk inside $\fp$ and boundary $\partial\fp$. It is one of the differences between the original proposal of exact holographic mapping in \cite{Qi2,Qi1} and the one emerged from LQG.

Indeed the state $|\Sig\rangle$ maps holographically a bulk state $\Phi_b$ to a boundary state $\Psi$, in the same way as Eq.\Ref{EHM0}. Here the bulk state $\Phi_b$ is generally a sum of the tenor products of bulk intertwiners $I_v$ at all vertices. For a suitable choice of $\Phi_b$\footnote{$\Phi_b$ used here is a sum of (coarse grained) tensor product states with equal weights. See Section \ref{EHM} for details.}, its image under holographic mapping $|\Sig\rangle$ gives a tensor network state $\Psi=\langle\Phi_b|\Sig\rangle$, representing the ground state of boundary CFT. The $\Phi_b$ relating to the tensor network is discussed in Sections \ref{EHM} and \ref{BSGC}. It exhibits the locality in the discrete space $\Sig$, i.e. its degrees of freedom are localized within each polyhedron $\fp$ and at each face $f$.

The fact that $|\Sig\rangle$ has the entanglement between bulk and boundary leads to some interesting consequences. The bulk state describes the quantum geometry inside $\Sig$, and the boundary state describes certain field theory on $\partial\Sig$. Then $|\Sig\rangle$ may be written schematically as an expansion in certain entangled basis
\be
|\Sig\rangle=\sum_{I}\big|\text{geometry}_I\big\rangle_b\otimes\big|\text{field}_I\big\rangle_\partial
\ee
where $I$ counts the bulk and boundary basis in $\ch_b$ and $\ch_\partial$. The entanglement in $|\Sig\rangle$ suggests the correspondence between boundary field and bulk geometry: If a measurement at the boundary gives $|\text{field}_I\rangle_\partial$, It makes the state collapses and determines the bulk state to be $|\text{geometry}_I\rangle_b$. Moreover, AdS/CFT correspondence suggests that if the boundary state is a CFT ground state $|\text{CFT}\rangle_\partial$, the corresponding bulk geometry state is a semiclassical state of AdS geometry. Thus $|\Sig\rangle$ is expected to have the form
\be
|\Sig\rangle=|\text{AdS}\rangle_b\otimes |\text{CFT}\rangle_\partial+\cdots \label{adscft0}
\ee
where $\cdots$ stands for other states in the expansion. From the viewpoint of holographic mapping, the CFT ground state is obtained by $|\text{CFT}\rangle_\partial={}_b\langle \text{AdS} |\Sig\rangle$. If we represent the $|\text{CFT}\rangle_\partial$ by the tensor network state $\Psi$, then the bulk state $\Phi_b$ proposed above should represent a semiclassical state of AdS. The semiclassicality is consistent with the locality exhibited by $\Phi_b$.  

The requirement that $\Phi_b$ represents the semiclassical geometry imposes constraints to $\Phi_b$. In particular, the geometry of $\Sig$ endows face areas $\Ar_f$ to each polyhedron faces $f$. Thus $\Phi_b$, as a sum of spin-networks, satisfies the constraints
\be
8\pi\g\ell_P^2\sum^{N_\G(f)}_{l,l\cap f\neq\emptyset}\sqrt{j_l(j_l+1)}=\Ar_f, \label{coarse0}
\ee
recall that $8\pi\g\ell_P^2\sqrt{j_l(j_l+1)}$ is the quantum area carried by a spin-network edge $l$. $N_\G(f)$ is the number of intersections between the spin-network graph and $f$. Eq.\Ref{coarse0} constraints the spin-network sum in $\Phi_b$, then effectively constrains the bond dimension\footnote{The bond dimension $D_f$ is the range of the index sum, when a pair of indices are contracted between $T(\fp)$ and $T(\fp')$ with $f=\fp\cap\fp'$.} $D_f$ of the tensor network $\Psi$. $D_f$ is constrained to be the number of microstates $\otimes_l|j_l,m_l\rangle$ at $f$ subject to Eq.\Ref{coarse0}. As $\Ar_f\gg \ell_P^2$, $D_f$ can be estimated by using the same technique as the LQG entropy counting on a black hole horizon, see e.g. \cite{GP2011,Ghosh:2004wq,QGandBH}. As a result, $\ln D_f$ is proportional to the face area $\Ar_f$.
\be
D_f\simeq \exp\lt[\mathrm{T}\,\Ar_{f}\rt]\gg 1,\label{arealaw0}
\ee
where $\mathrm{T}=\frac{\b_0}{8\pi\g\ell_P^2}$. $\g$ is the Barbero-Immirzi paramter in LQG, and $2\pi\b_0=0.274...$. This area law of $D_f$ is an important ingredient in the derivation of holographic entanglement entropy.

To show the exact holographic mapping $|\Sig\rangle$ from LQG indeed consistent with AdS/CFT, we specify a region $A$ and its complement $\bA$ in $\partial\Sig$, and compute the entanglement entropy $S(A)$ of the resulting tensor network $\Psi$ (as a candidate of $|\text{CFT}\rangle_\partial$). Thanks to the area law of bond dimension, in the semiclassical regime Eq.\Ref{lAL}, $S(A)$ reduces to a path integral of Nambu-Goto action $\mathrm{T}\,\Ar_{\cs}$ of the bulk 2-surfaces $\cs$ attached to $\partial A$ on the boundary, i.e. $\partial\cs=\partial A$. 
\be
e^{-S(A)}\simeq\int [D \cs]\, e^{- \text{T} \Ar_\cs}\label{NGaction0}
\ee
where $D\cs$ is certain measure of the surfaces $\cs$. In the limit $\ell_P\to 0$, the path integral localizes at the critical point, which is the surface $\cs$ with minimal area $\Ar_{\text{min}}$. As a result, we reproduces the Ryu-Takayanagi formula
\be
S(A)\simeq\mathrm{T}\,\Ar_{\text{min}}.\label{RT0}
\ee
where the ``surface tension'' $\mathrm{T}$ is identified as the IR value of $1/4G_{N}$\footnote{This is consistent with what has been suggested in the LQG black hole literature \cite{Ghosh:2012wq}.}.


In the derivation of holographic entanglement entropy, we use the similar technique as in \cite{Qi1}, i.e. we apply a random sampling of the state $|\cv_\fp\rangle$ at each $\fp$. The random state technique has a long and rich history in quantum information theory (see e.g \cite{randomrev} for a review), and has been often used in the studies of entanglement entropy (e.g. \cite{hayden,Page:1993df}). It is also motivated by the fact that each polyhedron $\fp$ contains a large number of spin-network microstates, whose linear combination gives $|\cv_\fp\rangle$. We are not interested in the detailed microstates at the Planck scale, but rather interested in a macroscopic result, e.g. the entanglement entropy, determined by the typical coarse grained state $|\cv_\fp\rangle$. The holographic entanglement entropy Eq.\Ref{RT0} is derived from the typical states $|\cv_\fp\rangle$ via random average. Any deviation from the above typical result can be shown to be suppressed as the bond dimension being large $D_f\gg1$. 

Although the idea of random sampling used here is similar to \cite{Qi1}, there are some important differences and improvements: Firstly we have pointed out that the state $|\cv_\fp\rangle$ from LQG is always an entangled state. The Hilbert space of $|\cv_\fp\rangle$ cannot be factorized into bulk of $\fp$ and boundary $\partial\fp$. So we have to impose this entanglement into the random sampling, and always sample the entangled states. It leads to a few technical differences between the derivation in Section \ref{EE} and in \cite{Qi1}. Because of this entanglement in $|\cv_\fp\rangle$, the bulk state $\Phi_b$, which leads to tensor network in holographic mapping, is not necessarily a pure tenor product, different from \cite{Qi1}, although it still represents the locality at both $\fp$'s and $f$'s.

Secondly, as a key input from LQG to the derivation, the bond dimension $D_f$ obtains the geometrical interpretation via the area law Eq.\Ref{arealaw0}, which is a key ingredient of writing $S(A)$ as a Nambu-Goto path integral and leads to the minimal surface by the variational principle. 

Although we often regard the tensor network as the CFT ground state, whose holographic dual is bulk AdS geometry, our formalism and the resulting Ryu-Takayanagi formula should be valid for the broader context of bulk-boundary duality. The only essential assumptions in the derivation are that the boundary quantum state should be a tensor network (may or may not be a CFT ground state), while the bulk quantum state should represent a bulk geometry (may or may not be AdS) with semiclassical areas, and represent bulk locality.   

In this paper we mainly focus on 4 spacetime dimension thus the space dimension $\dim\Sig=3$. All results generalize straight-forwardly to 3d spacetime when the surface areas $\Ar$ here are replaced by the lengths in 2d space. The generalization to spacetime dimensions higher than 4 can also be done by applying the state counting in \cite{Bodendorfer:2013sja} to estimate $D_f$. 

Back to the exact holographic mapping in the expansion Eq.\Ref{adscft0}, it seems to give a holographic dictionary between bulk and boundary quantum states, through their entanglement. We may also re-expand $|\Sig\rangle$ by another choice of bulk and boundary basis, which is slightly deviate from the ones containing $|\text{AdS}\rangle_b$ and $|\text{CFT}\rangle_\partial$. If the new bulk basis contains $|\text{AdS}+\text{perturb.}\rangle_b$, then the boundary state entangled with it should be $|\text{CFT}+\text{perturb.}\rangle_\partial$, i.e. the re-expansion of $|\Sig\rangle$ should lead to
\be
|\Sig\rangle=\lt|\text{AdS}+\text{perturb.}\rt\rangle_b\otimes \lt|\text{CFT}+\text{perturb.}\rt\rangle_\partial+\cdots
\ee
Thus certain excitations on the AdS geometry is entangled with certain excitations on the CFT ground state. So it might give a holographic dictionary between bulk and boundary operators. The concrete understanding of it is a research currently undergoing \cite{future}.

Let us mention that the present work mainly focuses on the kinematics of LQG and tensor network. The tensor construction thus far attacks mostly the kinematical aspect of the problem -- identifying wavefunctions that exhibit properties of semi-classical gravitational backgrounds. The result of holographic entanglement entropy (Ryu-Takayanagi formula) indeed supports that the tensor network state emerges from LQG is the ground state of certain CFT. The dynamics of this CFT is unclear at the moment, but interestingly, it should be possible to translate between dynamics of the CFT and dynamics of the quantum gravity. One important piece of the puzzle is how unitary transformations of the boundary legs can be translated into unitary transformations of the bulk tensors and under what circumstances such evolution would preserve the tensor network structure in the bulk. These might relate to the Hamiltonian constraint and Spinfoams in LQG. 

Finally we mention that there has been earlier works in the literature on the relation between LQG and holography, e.g. \cite{Markopoulou:1999iq,Ling:2000ss,Ghosh:2013iwa,Han:2014xna,Ghosh:2014rra,Bianchi:2012ev,Zuo:2016ezr,Bonzom:2015ans,hanSUSY,Yang:2008th,Freidel:2015gpa}, including a recent work on holographic entanglement entropy by Lee Smolin \cite{Smolin:2016edy}. There also has been works on the entanglement entropy in LQG context e.g. \cite{Perez:2014ura,Bianchi:2015fra,Bianchi:2014bma,Bodendorfer:2014fua,Delcamp:2016eya,Livine:2005mw,Donnelly:2016auv,Donnelly:2008vx,Donnelly:2011hn}, and on applying tensor network technique to LQG e.g. \cite{Dittrich:2014mxa}. Some earlier works on coarse graining spin-network states can be found in e.g. \cite{Dittrich:2016tys,Anza:2016fix,Dittrich:2013bza,Bodendorfer:2016tky,Livine:2013gna,Bahr:2012qj,Livine:2006xk}.

The architecture of this paper is as follows: Section \ref{LQGreview} reviews the Hilbert space of LQG and spin-network states. Section \ref{SNwB} focuses on the spin-network with boundary dangling edges, and presents a useful reformulation. Section \ref{CGSN} discusses the coarse graining of spin-network state, and define the state $|\cv_\fp\rangle$. Section \ref{EHM} define the LQG state $|\Sig\rangle$ and demonstrates the exact holographic mapping. Section \ref{BBE} discusses the bulk-boundary entanglement in $|\Sig\rangle$ and its relation with bulk-boundary duality. Section \ref{BSGC} discusses the bulk state $\Phi_b$ representing a semiclassical geometry with locality, and the geometrical constraints imposed to the bulk state. Section \ref{bond} estimates the bond dimension $D_f$ and shows the area law. Finally, Section \ref{EE} derives the Ryu-Takayanagi formula of holographic entanglement entropy.

\section{Hilbert space of Loop Quantum Gravity and Spin-Network}\label{LQGreview}

The Hilbert space of LQG is derived from quantizing the phase space $\cp$ of 4d gravity in the formulation using connection variables \cite{Ashtekar:1986yd,review,review1,book}. The phase space $\cp$ can be obtained via a 3+1 decomposition of the Holst action \cite{holst}
\be
S_{\text{Holst}}\lt[e^I,\o^{IJ}\rt]=\frac{1}{8\pi G_{N}}\int_{M_4} e^I\wedge e^J\wedge \lt(*F+\frac{1}{\g}F\rt)_{IJ}
\ee 
where $e^I_\mu$ is the 4d tetrad and $F^{IJ}_{\mu\nu}$ is the curvature of so(1,3) connection $\o^{IJ}_\mu$ ($\mu,I=0,\cdots,3$). $\g\in\R$ is the Barbero-Immirzi parameter. In this paper $\g$ is an arbitrary number of order 1. The variational principle of $S_{\text{Holst}}$ gives vacuum Einstein equation, and shows $S_{\text{Holst}}$ is on-shell equivalent to the Einstein-Hilbert action of gravity.  

The 3+1 decomposition and Hamiltonian analysis of $S_{\text{Holst}}$ leads to the phase space $\cp$ of 4d gravity. The canonical conjugate variables in $\cp$ are the Ashtekar-Barbero connection $A_a^i$ and densitized triad $E^a_i$ on the spatial slices $M_3$:
\be
A_a^i=\G_a^i+\g K^i_a,\quad E^a_i=\sqrt{\det q}\, e^a_i
\ee
Here $e^a_i$ ($a,i=1,2,3$) is the triad on $M_3$, which determines the metric $q_{ab}=e^i_a e^i_b$ and the spin connection $\G_a^i$. $K_a^i$ relates to the extrinsic curvature $K_{ab}$ of $M_3\hookrightarrow M_4$ by $K_{ab}=K^i_{(a}e^i_{b)}$. In contrast to the so(1,3) connection $\o^{IJ}_\mu$ in 4d, $A_a^i$ is a spatial connection in 3d with gauge group SU(2). The breaking of the gauge group from the Lagrangian $S_{\text{Holst}}$ to the Hamiltonian formulation is due to the 3+1 decomposition of spacetime, together with an internal partial gauge fixing (usually called ``time gauge'' in the literature). The detailed derivation of the canonical conjugate variables can be found in e.g. \cite{holst,barbero}. The symplectic structure of the phase space $\cp$ gives the Poisson bracket
\be
\{A_a^i(x),\ E^b_j(x')\}=8\pi G_{N}\g\ \delta^b_a\delta^i_j\delta^{(3)}(x,x'). 
\ee

The quantization of phase space $\cp$ has been well-understood in LQG literature. See e.g. \cite{Rovelli1988,Ashtekar:1991kc,book}. The wave function of the theory can be understood as a function $\psi(A_a^i)$ of connection field on $M_3$. More precisely, the wave functions are functions $\psi$ of SU(2) holonomies $h_e(A)=P\exp\int_e A$ along a number of oriented edges (analytic curves) $e_1,\cdots,e_N\subset M_3$:
\be
\psi=\psi\lt(h_{e_1}(A),\cdots h_{e_N}(A)\rt).\label{cylindrical}
\ee 
The edges $e_1,\cdots,e_N$ form a graph (a network) $\G$. A general graph $\G$ consists a finite set of oriented edges denoted by $E(\G)$ and a set of vertices $V(\G)$. The vertices in $\G$ are the sources and targets of the edges $e\in E(\G)$. A wave function Eq.\Ref{cylindrical} is defined upon a choice of graph $\G$, and depends on only a finite number of degree of freedom. Thus $\psi$ in Eq.\Ref{cylindrical} is referred to as a \emph{cylindrical function} of LQG. Obviously, the full infinite number of degree of freedom of gravity is achieved by putting together all possible choices of $\G$\footnote{The LQG Hilbert space is the completion of the union of all possible cylindrical functions, modulo some equivalence relations. In simple language, the equivalence relations include the cylindrical functions of a small graph into the cylindrical functions of a larger graph. The integration of cylindrical functions and all operators have to respect to the equivalence relation, which is known as cylindrical consistency.}. All possible cylindrical function of the type Eq.\Ref{cylindrical} by considering all possible $\G$ form a Hilbert space of $L^2$-type,
\be
\ch_{LQG}=L^2(\bar{\ca}/\bar{\cg},\rmd\mu_{AL})
\ee
$\ch_{LQG}$ is the Hilbert space of LQG, and carrying the representation of quantum geometry on spatial manifold $M_3$. The configuration space $\bar{\ca}/\bar{\cg}$ is the space of all SU(2) connection fields (including certain non-smooth and distributional connection fields) over the spatial slices $M_3$, modulo the gauge transformations \cite{Marolf:1994cj}. $\rmd\mu_{AL}$ denotes the Ashtekar-Lewandowski measure on $\bar{\ca}/\bar{\cg}$ \cite{Ashtekar:1993wf}. Importantly, the LQG quantization of the gravity phase space $\cp$ is systematic and mathematically rigorous. The formalism is even \emph{unique}. It is proved in \cite{uniqueness,Fleischhack:2007mj} that $\ch_{LQG}$ is the unique representation of the quantization of $\cp$, provided that the theory is diffeomorphism invariant. 

There is a useful orthonormal basis in $\ch_{LQG}$ which is called \emph{spin-network} basis. The spin-network basis can be constructed by the following observation: Consider the simplest graph consisting only a single edge $e$, the associated cylindrical functions $\psi=\psi(h_e)$ is a function on SU(2) group. More precisely $\psi$ belongs to the space $L^2(\text{SU(2)},\rmd\mu_H)$ where $\rmd\mu_H$ is the Haar measure. An orthogonal basis in $L^2(\text{SU(2)},\rmd\mu_H)$ is given by the matrix elements $R^j_{mn}(h_e)$ of all SU(2) irreps labelled by the spins $j$
\be
R^j_{mn}(h_e)=\langle j,m|h_e|j,n\rangle.
\ee 
Thus $\psi(h_e)$ can be written as a linear combination of $R^j_{mn}(h_e)$: $\psi=\sum_{j=0}^\infty\sum_{m,n=-j}^jc^{j}_{mn} R^j_{mn}$. A general cylindrical function $\psi(h_{e_1},\cdots,h_{e_N})$ on a close graph $\G$ can be decomposed in the same way at each entry $h_{e_i}$. The basis to make the decomposition is a product of $R^{j_e}_{m_en_e}(h_e)$ over all $e\in\G$. However in order to preserve the gauge invariance at each vertex $v\in V(\G)$, an invariant tensor $I_v$ has to be inserted and contract the $m_e,n_e$ from the adjacent $e$'s. Therefore we have the (gauge invariant) spin-network basis $T_{\G,\{j_e\},\{I_v\}}$ for decomposing arbitrary $\psi(h_{e_1},\cdots,h_{e_N})$
\be
T_{\G,\{j_e\},\{I_v\}}\lt(h_{e_1},\cdots,h_{e_N}\rt)=\sum_{m_e,n_e}\prod_{e\in E(\G)} R^{j_e}_{m_en_e}(h_e)\prod_{v\in V(\G)} (I_v)^{\{j_e\}}_{\{{m}_e,{n}_e\}}.
\ee
where $I_v\in\mathrm{Inv}_{\Su}(\otimes_e V_{j_e})$ is an invariant tensor in SU(2) tensor representation $\otimes_e V_{j_e}$ with $j_e$ on adjacent edges. $I_v$ is often called an \emph{intertwiner} of SU(2).

For closed graphs, the spin-network state is defined as a triple $|\G,\{j_e\},\{I_v\}\rangle$ containing the graph $\G$, spins $j_e$ associated to all edges, and intertwiners $I_v$ associated to all vertices. The vertex in a closed graph is not uni-valent. The spin-network state relates to the spin-network function $T_{\G,\{j_e\},\{I_v\}}\lt(h_{e_1},\cdots,h_{e_N}\rt)$ by
\be
\lag h_{e_1},\cdots,h_{e_N}\big|\G,\{j_e\},\{I_v\}\rag=T_{\G,\{j_e\},\{I_v\}}\lt(h_{e_1},\cdots,h_{e_N}\rt).
\ee

A non-closed graph $\G$ with dangling edges $l$ (and uni-valent vertices) is used to define spin-network states for the spatial region $\calr$ with boundary $\partial\calr$. The corresponding spin-network state is denoted by $|\G,\{j_e,j_l\},\{I_v\},\{n_l\}\rangle$, where both a spin $j_l$ and a magnetic quantum number $n_l$ is assigned to each dangling edge $l$.  
\be
\lag \{h_{e}\},\{h_{l}\}\Big|\G,\{j_e,j_l\},\{I_v\},\{n_l\}\rag=\sum_{m_e,n_e,m_l}\prod_{e} R^{j_e}_{m_en_e}(h_e)\prod_{v} (I_v)^{\{j_e,j_l\}}_{\{{m}_e,{n}_e,m_l\}}\prod_lR^{j_l}_{m_ln_l}(h_l)\label{snfunc}
\ee
We often use $e$ to label internal edges and $l$ to label boundary dangling edges. It is clear from the above formula that the SU(2) gauge invariance is imposed on each internal vertex $v$, but is not imposed on the boundary uni-valent vertices. The gauge degree of freedom becomes the physical degree of freedom at the boundary. 

The spin-network basis is a complete orthogonal basis in the LQG Hilbert space $\ch$. It also has nice interpretations in terms of 3d quantum geometry on spatial manifold $M_3$. A large class of geometrical quantities on $M_3$ have been represented as the self-adjoint operators acting on $\ch$, constructed using the derivative $\hat{E}^a_i(x)=-8\pi i\g\ell_P^2\frac{\delta }{\delta A^i_a(x)}$. See e.g. \cite{Rovelli1995,ALarea,ALvolume,Thiemann:1996at,Bianchi:2008es,Ma:2010fy,Ma:2000au}. Among the class of geometrical operators, two important ones are the area operator $\hat{\Ar}_S$ of a 2-surface $S$, and the volume operator $\hat{V}_{\calr}$ of a 3d region $\calr$. It turns out that the spin-network basis is simultaneously the eigen-basis of both $\hat{\Ar}_S$ and $\hat{V}_{\calr}$. For $S$ cuts transversely a number edges in $\G$ and $\calr$ encloses a number of vertices
\be
\hat{\Ar}_S\Big|\G,\{j_e,j_l\},\{I_v\},\{n_l\}\Big\rangle&=&8\pi\g\ell_P^2\lt(\sum_{S\cap e\neq\emptyset}\sqrt{j_{e}(j_{e}+1)}+\sum_{S\cap l\neq\emptyset}\sqrt{j_{l}(j_{l}+1)}\rt) \Big|\G,\{j_e,j_l\},\{I_v\},\{n_l\}\Big\rangle\nonumber\\
\hat{V}_{\calr}\Big|\G,\{j_e,j_l\},\{I_v\},\{n_l\}\Big\rangle&=&\sum_{v\in\calr} V_v(I_v)\Big|\G,\{j_e,j_l\},\{I_v\},\{n_l\}\Big\rangle
\ee
The eigenvalue of area operator is given by a sum over all cut edges weighted by $\sqrt{j(j+1)}$ according to the spin labels carried by the edges. The spin labels $j_e,j_l$ is thus interpreted as the quanta of the area element transverse to the edge. The eigenvalue of volume operator is a sum of local contribution $V_v(I_v)$ at $v\in\calr$, which is determined by the intertwiner $I_v$ at $v$ \cite{Brunnemann:2007ca,Bianchi:2011ub}. One may imagine that there is a tiny flat polyhedron\footnote{This tiny polyhedron is used for the geometrical picture of spin-network vertex $v$. It shouldn't be confused with the semiclassical polyhedron $\fp$ in Section \ref{CGSN} and the following. The semiclassical polyhedron $\fp$ contains a large number of spin-network vertices, thus can be obtained by certain gluing of a large number of tiny polyhedra.}  (infinitesimal from macroscopic point of view) enclosing a single vertex $v$. The polyhedron faces are all transverse to the edges adjacent to $v$. The intertwiner $I_v$ is a quantization of the shapes of the tiny polyhedron \cite{shape,CF}, while the $j$'s on the edges adjacent to $v$ gives the quantum areas of the tiny polyhedron faces.


\section{Spin-Network with Boundary}\label{SNwB}

In this section, we focus on the spin-networks with dangling edges, describing the quantum geometry of the spatial region $\calr$ with boundary $\partial\calr$. We present a different formulation of the spin-network states, instead of using spin-network functions Eq.\Ref{snfunc}. This formulation is useful in the following discussion.

Given an oriented graph $\G$, we cut each edge $e\in E(\G)$ into a pair of half-edges. The cuts break $\G$ into a set of vertices $v\in V(\G)$, together with the adjacent half-edges $e_{1,\cdots,n}(v)$. Some of the half-edges are incoming to $v$, while others are outgoing. Now we focus on a single vertex $v$, and associates each half-edge $e_i(v)$ a Hilbert space $L^{2}(\Su,\rmd\mu_H)\simeq \oplus_{j=0}^\infty V_j\otimes V_j^*$ where $V_j$ is the spin-$j$ irrep of SU(2). The factorization naturally associates $V_j$ to the source $s(e_i(v))$ of $e_i(v)$, and associates $V_j^*$ to the target $t(e_i(v))$. Thus the vertex $x$ associates a product of irreps $\otimes_i V_{j_i}\otimes_l V_{j_l}^*$, where $V_{j_i}, V_{j_l}^*$ corresponds to the outgoing and incoming half-edges respectively. SU(2) gauge invariance at each vertex restricts the product of irreps into the invariant subspace $\mathrm{Inv}[\otimes_i V_{j_i}\otimes_l V_{j_l}^*]$ (intertwiner space), where each state is invariant under the tensor product representation of SU(2). Combining the other factor $V_j$ or $V_j^*$ half-edge Hilbert space $L^{2}(\Su)$, this construction associates a Hilbert space to each vertex $x$ and its adjacent half-edges
\be
\ch_v= \bigoplus_{\vec{j}} \mathrm{Inv}\lt[\bigotimes_{e_i\ \text{outgoing}} V_{j_i}\bigotimes_{e_l\ \text{incoming}} V_{j_l}^*\rt]\bigotimes\lt[\bigotimes_{e_i\ \text{outgoing}} V^*_{j_i}\bigotimes_{e_l\ \text{incoming}} V_{j_l}\rt].
\ee 

Given a basis $|I^{\vec{j}}_v\rangle= \sum_{\vec{m}}\otimes_{e_i}|j_i,m_i\rangle\, I^{\vec{j}}_{\vec{m}} \in \mathrm{Inv}[\bigotimes_{e_i} V_{j_i}]$ (all $e_i$ are assumed outgoing), a basis in $\ch_v$ can be written as
\be
\lt(\sum_{\vec{m}}\otimes_{e_i}|j_i,m_i\rangle\, I^{\vec{j}}_{\vec{m}}\rt)\otimes_{e_i}\langle j_i, n_i|=|I^{\vec{j}}_v\rangle\otimes_{e_i}\langle j_i, n_i|\label{V}
\ee
We have assumed all half-edges $e_i$ are outgoing from the vertex $x$ to simplify the formula. The most general case with both incoming and outgoing $e_i$'s can be written analogously.

A LQG spin-network state $|\G,\{j_e,j_l\},\{I_v\},\{n_l\}\rangle$ on a graph $\G$ with boundary dangling edges is obtained by taking the inner product between pairs of $\langle j_i, n_i|$ for each internal edge. 
\be
\Big|\G,\{j_e,j_l\},\{I_v\},\{n_l\}\Big\rangle=\bigotimes_{v\in V(\G)} \lt|I^{\{j_e,j_l\}}_v\rag\bigotimes_{\text{boundary}\ l}\langle j_l, n_l|.\label{SNboundary}
\ee
The last tensor product is among the dangling half-edges $l$ connecting to $\partial\calr$, which do not participate the inner product. The corresponding spin-network function for each $|\G,\{j_e,j_l\},\{I_v\},\{n_l\}\rangle$ is
\be
\lag h_e,h_l \Big|\G,\{j_e,j_l\},\{I_v\},\{n_l\} \rag=\sum_{m_e,n_e,m_l}\prod_e R^{j_e}_{m_en_e}(h_e)\prod_x I^{\{j_e,j_l\}}_{\{{m}_e,{n}_e,m_l\}}(x)\prod_lR^{j_l}_{m_ln_l}(h_l)
\ee
To derive the above expression, one may notice that $|\G,\{j_e,j_l\},\{I_v\},\{n_l\}\rangle$ contains for each edge $e$ or $l$ a state $|\phi^j_{nm}\rangle = |j,m\rangle\langle j,n|$, where one of the factor comes from $I^{\{j_e\}}_v$ and the other comes from either a neighboring $I^{\{j_e\}}_{v'}$ or a dangling half-edge $l$. We have $\langle h|\phi^j_{nm}\rangle= \langle j,n| h |j,m\rangle=R^j_{nm}(h)$.

\section{Coarse Grained Spin-Network}\label{CGSN}

Any 3d spatial manifold can be discretized into a (large) number of polyhedra. Given a polyhedral region denoted by $\fp$, we assume that the scale of polyhedron, given by the geometry endowed on $\fp$, is much larger than the Planck scale, or the scale of spin-network quantum geometry. Therefore the geometry of $\fp$ should be described by the spin-networks with a large number of edges and vertices. We consider a generic LQG state $|\cv_\fp\rangle\in\ch_\fp$ with boundary in $\fp$
\be
|\cv_\fp\rangle&=&\sum_{\G_\fp,\{j_e,j_l\},\{I_v\},\{n_l\}}\cv_{\G_\fp,\{j_e,j_l\},\{I_v\},\{n_l\}}\big|\G_\fp,\{j_e,j_l\},\{I_v\},\{n_l\}\big\rangle\nonumber\\
&=&\sum_{\G_\fp,\{j_e,j_l\},\{I_v\},\{n_l\}}\cv_{\G_\fp,\{j_e,j_l\},\{I_v\},\{n_l\}}\bigotimes_{v\in V(\G_\fp)} \lt|I^{\{j_e,j_l\}}_v\rag\bigotimes_{\text{boundary}\ l}\langle j_l, n_l|.\label{cvp}
\ee
where we have summed over all graphs $\G_\fp$ in the region $\fp$. Notice that for a given graph $\G_\fp$, there are a number of boundary edges $l$ intersecting the 2d boundary of polyhedron. For simplicity, we only consider in the above sum the graphs $\G_\fp$ having intersection with the polyhedron faces, while discarding the case of intersection on polyhedron edges or vertices. The Hilbert space $\ch_\fp$ can be expressed using SU(2) irreps $V_j$: 
\be
\ch_\fp=\bigoplus_{\G_\fp}\bigoplus_{\{j_e,j_l\}}\bigotimes_{v\in V(\G_\fp)}\mathrm{Inv}\lt[\bigotimes_{e\ \text{outgoing}} V_{j_e}\bigotimes_{e\ \text{incoming}} V_{j_e}^*\rt]\bigotimes_{l,l\cap f\neq\emptyset} V_{j_l}.\label{chp}
\ee

We will consider the 3d spatial manifold carrying a semi-classical geometry, which endows a polyhedral geometry to each $\fp$. The geometrical data of each $\fp$ include in particular the areas of all its faces $\Ar_f$, as well as other geometrical quantities e.g. curvature, shape, etc. It is important that, here we assume the scale of this geometrical polyhedron $\fp$ is much large than the Planck scale, where the geometry is quantized by spin-networks. Recall that the spin-network states present the 3d quantum geometry at Planck scale. Fundamentally, the complete quantum geometry of a relatively large polyhedron needs the spin-network states associated to arbitrary graph $\G_\fp$ (with a large number of edges and vertices) in the polyhedron. The polyhedron $\fp$ being semi-classical implies that $\Ar_f\gg\ell_P^2$. So there may be a large number of dangling spin-network edges intersecting each polyhedron face $f\subset\partial\fp$, while each spin-network edge carries a quantum area of the same order as $\ell_P^2$. The bulk degrees of freedom in $\fp$ is given by the spins $j_e$ on the internal edges and the intertwiners $I_v$ at the (internal) vertices. They are the quantum face areas and quantum shapes of the Planck scale polyhedra located at the vertices $v$. The semiclassical geometries of the polyhedron $\fp$ are obtained by gluing a large number of these tiny geometrical polyhedra. 

It is thus clear that within a semiclassical polyhedron $\fp$, there is a large number of quantum geometry micro-degrees of freedom, which in principle should be described by the spin-network states on a large graph $\G_\fp$. However, here we would like to find an effective prescription for the states of type $|\cv_\fp\rangle$, by smearing the detailed structure of $|\cv_\fp\rangle$ inside $\fp$. The effective prescription is a coarse graining of the micro-degrees of freedom in $\fp$, as an analog of the block spin procedure in Ising model. 

For any spin-network state $\big|\G_\fp,\{j_e,j_l\},\{I_v\},\{n_l\}\big\rangle$, given a face $f$ of $\fp$ with $N_{\G}(f)$ intersections, the total area of polyhedron face $f$ is given by 
\be
8\pi\g\ell_P^2\sum^{N_\G(f)}_{l,l\cap f\neq\emptyset}\sqrt{j_l(j_l+1)}=\Ar_f.
\ee
We would like to make the state $|\cv_\fp\rangle$ as a superposition of all possible quantum geometry of the polyhedron, including all generic quantum fluctuations. Thus $|\cv_\fp\rangle$ defined in Eq.\Ref{cvp} is a sum over spin-networks. We do not fix the value of total area $\Ar_f$ for $|\cv_\fp\rangle$, and let the sum over $j_l$ in Eq.\Ref{cvp} unconstrained. So $|\cv_\fp\rangle$ also contains a superposition of the semiclassical polyhedron geometries with different face areas.

A Hilbert space $\ch_\partial(f)$ associated to the face $f$ can be defined to be
\be
\ch_\partial(f)=\bigoplus_{N_\G(f)}\bigoplus_{\{j_l\}}\bigotimes^{N_\G(f)}_{l,l\cap f\neq\emptyset} V_{j_l}. \label{chf}
\ee 
We may make this Hilbert space finite-dimensional by imposing a cut-off $\Ar_f\leq \L^{-1}$ where $\L$ is the cosmological constant, i.e. the scale of the polyhedron $\fp$ shouldn't exceed the cosmological scale. Then the range of each $j_l$ becomes finite $j_l\leq ({8\pi\g\ell_P^2\L})^{-1}$ \footnote{In LQG the cosmological constant is implemented by using quantum group or Chern-Simons theory \cite{QSF,QSF1,3dblockHHKR,HHKRshort,HHKR}}. The first direct sum over all possible intersection numbers $\oplus_{N_\G(f)}$ also becomes a finite sum by $N_\G(f)\leq ({4\pi\g\ell_P^2\sqrt{3}\L})^{-1}$. The maximal $N_\G(f)$ is obtained by assuming all lowest $j_l=1/2$ and saturating $\L^{-1}$.


We define another Hilbert space $\ch_b(\fp)$ of the states in the bulk of $\fp$.
\be
\ch_b({\fp})=\bigoplus_{\G_\fp}\bigoplus_{\{j_e\}}\bigotimes_{v\in V(\G_\fp)}\mathrm{Inv}\lt[\bigotimes_{e\ \text{outgoing}} V_{j_e}\bigotimes_{e\ \text{incoming}} V_{j_e}^*\rt]\label{Hbp}
\ee
The labels $e$ in the above formula also include the boundary edges $l$. It is easy to see that a LQG state $|\cv_\fp\rangle$ is a linear map between $\ch_b(\fp)$ and $\otimes_f \ch_\partial(f)$
\be
|\cv_\fp\rangle:\ \ch_b(\fp)\rightarrow\bigotimes_f \ch_\partial(f),
\ee
although the Hilbert space $\ch_\fp$ of $|\cv_\fp\rangle$ is only a subspace of the tensor product $\ch_b(\fp)\otimes_f \ch_\partial(f)$ because of the matching of $j_l$ in Eq.\Ref{cvp}, between the close-to-boundary $|I_v^{j_e,j_l}\rangle$ and the boundary $|j_l,n_l\rangle$. The state $|\cv_\fp\rangle$ may be viewed as a simple example of exact holographic mapping from bulk states to boundary states. This point is further explored in the next section. 

To simplify the notation, we denote the coarse grained basis in $\ch_\partial(f)$ and $\ch_b(\fp)$ by 
\be
|\mu_f\rangle\equiv \bigotimes_{l\cap f\neq\emptyset}\langle j_l, n_l|,\quad |\xi^{\vec{\mu}}_\fp \rangle\equiv \bigotimes_{v\in V(\G_\fp)} \lt|I^{\{j_e,j_l\}}_v\rag
\ee
$|\mu_f\rangle$ and $|\xi^{\vec{\mu}}_\fp \rangle$ label the boundary and bulk microstates of the polyhedron $\fp$. $|\mu_f\rangle$ characterizes the degrees of freedom of surface geometry on $f$ \cite{polymer}, while $|\xi^{\vec{\mu}}_\fp \rangle$ characterize the degrees of freedom of bulk geometry in $\fp$. 
However it turns out that the geometrical information encoded by the bulk states $|\xi^{\vec{\mu}}_\fp \rangle$ doesn't play much role in the present analysis in Sections \ref{2nd} and \ref{higher}, in which we focus on the area law (Ryu-Takayanagi formula) of the holographic entanglement entropy. 
 
By using the new notation, the state $|\cv_\fp\rangle$ is now written as
\be
|\cv_\fp\rangle=\sum_{\mu_f,\xi_\fp} \cv_{\mu_f,\xi_\fp}|\xi^{\vec{\mu}}_\fp,\mu_f\rangle=\sum_{\mu_f,\xi_\fp} \cv_{\mu_f,\xi_\fp}|\xi^{\vec{\mu}}_\fp\rangle\otimes|\mu_f\rangle\label{coarsecv}
\ee
which exhibits certain entanglement between bulk and boundary states. As a linear combination of geometrical microstates $|\xi^{\vec{\mu}}_\fp\rangle\otimes|\mu_f\rangle$, the state $|\cv_\fp\rangle$ encodes certain quantum fluctuation of 3d geometry.

Given a state $\phi_b(\fp)\in \ch_b(\fp)$, its partial inner product with $|\cv_\fp\rangle$ produce a tensor state in $\otimes_f \ch_\partial(f)$
\be
|T(\fp)\rangle= \langle \phi_b(\fp)|\cv_\fp\rangle\in\bigotimes_f \ch_\partial(f)\label{tensor}
\ee 
For any given orthonormal basis $\{|\mu_f\rangle\}$ in $\ch_\partial(f)$ (there are $M$ polyhedron faces $f=1,\cdots,M$), the tensor $|T(\fp)\rangle$ can be expressed as
\be
|T(\fp)\rangle=\sum_{\mu_{1},\cdots,\mu_{M}}T(\fp)_{\mu_{1},\cdots,\mu_{M}}|\mu_1\rangle\otimes\cdots\otimes|\mu_M\rangle.
\ee
where $\mu_f=1,\cdots,\dim\ch_\partial(f)$. The above inner product with bulk state $\phi_b(\fp)$ effectively describes a procedure of integrating out the bulk degrees of freedom, which gives the tensor $|T(\fp)\rangle$ as the effective boundary state.

\section{Exact Holographic Mapping and Tensor Network}\label{EHM}

We consider a (large) spatial region $\Sig$ with nontrivial boundary $\partial\Sig$. $\Sig$ can be decomposed by a large number of semiclassical polyhedra $\fp$. All detailed quantum geometries of polyhedra are stored in the states $|\cv_\fp\rangle$ for all $\fp$. The averaged length scale of each semiclassical polyhedra $\fp$ is assumed to be small comparing to the macroscopic scale, although it is large comparing to the Planck scale. The (squared) length scale of $\fp$ is represented by the mean face area $\Ar_f$, while the macroscopic scale is represented by e.g. the mean curvature radius $L$ of the macroscopic geometry built by gluing $\fp$'s. The analysis in this paper mainly focus on the regime that
\be
\ell_P^2\ll \Ar_f\ll L^2.
\ee
This regime has been studied as the semiclassical regime of LQG from several different perspectives \cite{Ghosh:2013iwa,hanBH,Han:2016fgh,lowE,Sahlmann:2001nv}. 

Therefore it is clear that we should consider the gluing of a large number polyhedra $\fp$ to obtain $\Sig$. Each of the polyhedra $\fp$ carries the quantum state $|\cv_\fp\rangle$, packaging a large number of spin-network microstates. The resulting quantum state of $\Sig$, denoted by $|\Sig\rangle$ will be also a linear combination of the spin-network microstates, and reduce to $|\cv_\fp\rangle$ in each $\fp$.

We firstly describe the construction of the state for $\Sig$ at the level of coarse grained spin-networks. To glue a pair of polyhedra $\fp,\fp'$ through a common face $f=\fp\cap\fp'$, we first identify their total areas $\Ar_f$ and their Hilbert spaces $\ch_\partial(f)$, then define a state $| f\rangle$ in $\ch_\partial(f)\otimes \ch_\partial(f)$, where the two copies of $\ch_\partial(f)$ associates to $\fp$ and $\fp'$ respectively.
\be
|f\rangle=\sum_{\mu_f}|\mu_f\rangle\otimes|\mu_f\rangle
\ee 
It is not hard to see that the following partial inner product gives a state in the glued polyhedra $\fp\cup\fp'$:
\be
\langle f|\,\Big(|\cv_\fp\rangle\otimes |\cv_{\fp'}\rangle\Big).
\ee

It is straight-forward to generalize to the gluing of an arbitrary large number of polyhedra $\fp=1,\cdots,N$ $(N\gg1)$ to form a large spatial region $\Sig=\cup_\fp \fp$ with boundary $\partial\Sig$. We introduce $|E\rangle\equiv \prod_f|f\rangle$ for all gluing interfaces $f$. The resulting state is 
\be
|\Sig\rangle= \langle E|\,\otimes_\fp|\cv_\fp\rangle.
\ee

The state $|\Sig\rangle$ is a linear combination of spin-network states whose graph $\G$ is in $\Sig$ with dangling edges intersecting $\partial\Sig$. The inner product with $|E\rangle =\otimes_f |f\rangle$ identifies the surface states $|\mu_f\rangle$ at the interface between $|\cv_\fp\rangle$ and $|\cv_{\fp'}\rangle$. Recall $|\mu_f\rangle$ is a short-hand notion of $\otimes_l |j_l,m_l\rangle$ for a number $N$ of intersections $l\cap f\neq\emptyset$, where $N$, $\{j_l\}$, and $\{m_l\}$ is determined by $|\mu_f\rangle$. The inner product with $|E\rangle$ thus connects the $N$ pairs of edges intersecting $f$ from $|\cv_\fp\rangle$ and $|\cv_{\fp'}\rangle$, and identifies the $\{j_l\}$, and $\{m_l\}$ labels. The inner product projects out the case that $|\cv_\fp\rangle$ and $|\cv_{\fp'}\rangle$ give two different numbers $N\neq N'$ of intersections at each $f$. Because of the inner product, the $\bigotimes_{l}\langle j_l, n_l|$ piece of Eq.\Ref{cvp} are contracted between $|\cv_\fp\rangle$ and $|\cv_{\fp'}\rangle$. The resulting states is thus a linear combination of spin-networks as Eq.\Ref{SNboundary} defined on the entire graph $\G$, where $\G=\cup_\fp \G_\fp$ is obtained by connecting each piece of $\G_\fp$ through the intersections at $f\subset\partial\fp$. Therefore we always have that $|\Sig\rangle$ is a LQG state:
\be
|\Sig\rangle= \langle E|\,\otimes_\fp|\cv_\fp\rangle\in\ch_{LQG}.
\ee

The resulting state $|\Sig\rangle$ can be understood as a concrete realization of exact holographic mapping proposed in \cite{Qi1,Qi2}, and it now comes from $\ch_{LQG}$ derived from a systematic quantization procedure of gravity. Indeed, $|\Sig\rangle$ is a linear map from the bulk Hilbert space ${\ch}_b$ to the boundary Hilbert space $\ch_\partial$. ${\ch}_b$ and $\ch_\partial$ are defined as follows:
\be
{\ch}_b=\bigotimes_\fp\ch_{b}(\fp),\quad
\ch_\partial=\bigotimes_{f\subset\partial\Sig}\ch_\partial(f).\label{Hb}
\ee
Given a bulk state $\Phi_b\in \ch_b$, $|\Sig\rangle$ maps $\Phi_b$ to a boundary state $\Psi\in\ch_\partial$ by using the partial inner product:
\be
\langle\Phi_b|\Sig\rangle=\langle\Phi_b|\otimes\langle E|\otimes_\fp|\cv_\fp\rangle\equiv|\Psi\rangle.\label{PhiEV}
\ee
In this way, an exact holographic correspondence between bulk states and boundary states is made by $|\Sig\rangle$, and it is the reason why $|\Sig\rangle$ is referred to as an exact holographic mapping. In this context, a single $|\cv_\fp\rangle$ is the simplest exact holographic mapping with a single polyhedron $\fp$ in the bulk.

As a remark, ${\ch}_b$ is defined as a direct product in Eq.\Ref{Hb}. It contains the states $\otimes_\fp |\xi_\fp^{\vec{\mu}}\rangle$, in which $\mu_f,\mu_f'$ from two neighboring $\fp,\fp'$ doesn't match at $f=\fp\cap\fp'$. But these states belong to the kernel of the holographic mapping, i.e. $|\Sig\rangle$ projects out these states and only keeps the states $\otimes_\fp |\xi_\fp^{\vec{\mu}}\rangle$ in which $\mu_f$ matches from $\fp,\fp'$ at $f$. From the definition of $\ch_b(\fp)$ in Eq.\Ref{Hbp}, the states in ${\ch}_b$ with nontrivial images $|\Psi\rangle$ precisely spans the space of bulk spin-networks
\be
\bigoplus_{\G}\bigoplus_{\{j_e\}}\bigotimes_v\mathrm{Inv}\lt[\bigotimes_{e\ \text{outgoing}} V_{j_e}\bigotimes_{e\ \text{incoming}} V_{j_e}^*\rt].\label{bulkspace}
\ee
in which a pair of intertwiners share the same spin label if they are connected by an edge. Here $\G$ is a large spin-network graph obtained by connecting $\G_\fp$'s according to the gluing of polyhedra. The space of bulk spin-networks is spanned by the states $\bigotimes_{v\in V(\G)} |I^{\{j_e,j_l\}}_v\rangle$ being a product of intertwiners (with matching of spins along edges). The bulk spin-network is different from Eq.\Ref{SNboundary} up to the boundary pieces $\otimes_l\langle j_l,n_l|$. It is clear that the inner product between Eq.\Ref{SNboundary} and the bulk spin-network produces a boundary state in $\ch_\partial$.

Similar to $|\cv_\fp\rangle$ of polyhedron, $|\Sig\rangle$ only belongs to a subspace of the tensor product ${\ch}_b\otimes\ch_\partial$. In the subspace, the matching of $\mu_f$ is imposed between the close-to-boundary $|\xi_\fp^{\vec{\mu}}\rangle$ and the boundary $|\mu_f\rangle$, due to Eq.\Ref{coarsecv}. Therefore $|\Sig\rangle$ exhibits certain entanglement between bulk and boundary states in $\ch_b$ and $\ch_\partial$, similar as $|\cv_\fp\rangle$ shown in Eq.\Ref{coarsecv}. In this sense, $|\Sig\rangle$ from LQG is more restricted than the exact holographic mapping proposed in \cite{Qi1,Qi2}.


We consider a special class of bulk state $\Phi_b$, being a sum of pure tensor product states 
\be
|\Phi_b\rangle=\sum_{\{\mu_f\}}^{\{D_f\}}\otimes_{\fp}\lt|\phi(\fp)_{\vec{\mu}}\rt\rangle,\quad \lt|\phi(\fp)_{\vec{\mu}}\rt\rangle=\sum_{\xi_\fp}\phi(\fp)_{\xi_\fp,\vec{\mu}}|\xi_\fp^{\vec{\mu}}\rangle.\label{cla}
\ee
Each $|\phi(\fp)_{\vec{\mu}}\rangle$ has been assumed to be normalized for any $\vec{\mu}$. The coefficients in the sum $\sum_{\{\mu_f\}}$ has been assumed to be factorized into local contributions $\phi(\fp)_{\xi_\fp,\vec{\mu}}$ \footnote{A more general $\Phi_b$ might be $
|\Phi_b\rangle=\sum_{\xi_\fp,\vec{\mu}}\Phi_{\vec{\xi},\vec{\mu}}\otimes_\fp|\xi_\fp^{\vec{\mu}}\rangle$, whose coefficients doesn't factorizes.}. $\Phi_b$ is a sum of tensor product states with equal weights. The matching of $\mu_f$ at each interface $f$ between $\fp,\fp'$ has been imposed, so that $|\Phi_b\rangle$ doesn't belong to the kernel of the exact holographic mapping $|\Sig\rangle$. We have assumed the range of each $\mu_f$ in the sum is constrained by a cut-off $D_f$, whose physical meaning is clear in Section \ref{BSGC}. 

Applying the exact holographic mapping $|\Sig\rangle$ to the bulk state $\Phi_b$ of this type, the corresponding boundary state $\Psi$ is a tensor network state
\be
|\Psi\rangle=\sum_{\{\mu_f\}}^{\{D_f\}}\prod_\fp T(\fp)_{\{\mu_f\}_{f\subset\partial\fp}}\otimes_{f\subset\partial\Sig}|\mu_f\rangle.\label{TNork}
\ee
Here each tensor $T(\fp)$ is associated to a polyhedron $\fp$, as defined in Eq.\Ref{tensor}. The bond dimensions of the tensor network are the ranges of the contracted indices $\mu_f$ between tensors, which is $D_f$. Being the bulk state relating to the tensor network, $\Phi_b$ in Eq.\Ref{cla} generalizes from the pure tensor product bulk state in \cite{Qi1} to a equal-weight sum of tensor products. 

The tensor network state is understood as a possible prescription of the ground state of CFT living on the boundary $\partial\Sig$, see e.g.\cite{Pastawski:2015qua,Almheiri:2014lwa}. Thus $\Phi_b$ of the type in Eq.\Ref{cla} stands out as far as we focus on the boundary CFT ground state. Moreover $\Phi_b$ satisfying Eq.\Ref{cla} represents the bulk locality. All the degrees of freedom of $\Phi_b$ are indeed local, because each factor $|\phi(\fp)_{\vec{\mu}}\rangle$ is localized within a polyhedron $\fp$, while each summed label $\mu_f$ is localized at the face of $\fp$. In Sections \ref{2nd} and \ref{higher}, the main result of the holographic entanglement entropy is derived for the boundary CFT ground state represented by tensor network. The main derivation is based on the bulk state satisfying Eq.\Ref{cla}.

\section{Bulk-Boundary Entanglement}\label{BBE}

It has been discussed that the exact holographic mapping $|\Sig\rangle$ from LQG exhibits the entanglement between bulk and boundary states in $\ch_b$ and $\ch_\partial$. If the bulk states in $\ch_b$ describes the quantum geometry living in $d$ dimensions, while the boundary states in $\ch_\partial$ describe the field theory living in $(d-1)$-dimensional boundary, $|\Sig\rangle$ may be written schematically as
\be
|\Sig\rangle=\sum_{I}\big|\text{geometry}_I\big\rangle_b\otimes\big|\text{field}_I\big\rangle_\partial
\ee
where $I$ labels the orthonormal basis in both Hilbert spaces. AdS/CFT correspondence suggests that if the boundary field theory state is a CFT ground state $|\text{CFT}\rangle$, the entangled bulk geometry should be a semiclassical geometry state $|\text{AdS}\rangle$, i.e.
\be
|\Sig\rangle=|\text{AdS}\rangle_b\otimes |\text{CFT}\rangle_\partial+\cdots\label{adscft}
\ee 
where $\cdots$ stands for the contribution from other bulk quantum geometries entangled with other boundary field theory states. If the quantum measurement is performed at the boundary whose output exhibits a CFT ground state, then $|\Sig\rangle$ collapses to the first term in the above expansion, which determines the bulk state to be a semiclassical AdS geometry. 

Interestingly, a holographic dictionary might be extracted from the above expansion. Since the above expansion depends on a choice of basis in both $\ch_b$ and $\ch_\partial$, we can re-expand $|\Sig\rangle$ by another choice of orthonormal basis in both $\ch_b$ and $\ch_\partial$, which is slightly deviate from the original basis containing $|\text{AdS}\rangle_b$ and $|\text{CFT}\rangle_\partial$. If the new basis in $\ch_b$ contains $|\text{AdS}+\text{perturb.}\rangle_b$, then the boundary state entangled with it should be $|\text{CFT}+\text{perturb.}\rangle_\partial$, i.e.
\be
|\Sig\rangle=\lt|\text{AdS}+\text{perturb.}\rt\rangle_b\otimes \lt|\text{CFT}+\text{perturb.}\rt\rangle_\partial+\cdots
\ee
Thus certain excitations on the AdS geometry is entangled with certain excitations on the CFT ground state. This entanglement might be a representation of the holographic dictionary of AdS/CFT. The concrete understanding of the entangled excitations is a research undergoing \cite{future}.

\section{Bulk State and Geometrical Constraint}\label{BSGC}

From the scheme in Eq.\Ref{adscft}, the CFT ground state $|\text{CFT}\rangle_\partial$ is extracted by the inner product ${}_b\langle\text{AdS}|\Sig\rangle$, which manifests $|\Sig\rangle$ to be an exact holographic mapping. So if $|\text{CFT}\rangle_\partial$ is represented by the tensor network state $\Psi$ in Eq.\Ref{TNork}, then $\Phi_b$ satisfying Eq.\Ref{cla} should represent the semiclassical AdS geometry in the bulk. 

The state $\Phi_b$ represents the semiclassical geometry and locality. Recall $|\Phi_b\rangle=\sum_{\{\mu_f\}}^{\{D_f\}}\otimes_{\fp}\lt|\phi(\fp)_{\vec{\mu}}\rt\rangle$ where $\lt|\phi(\fp)_{\vec{\mu}}\rt\rangle=\sum_{\xi_\fp}\phi(\fp)_{\xi_\fp,\vec{\mu}}|\xi_\fp^{\vec{\mu}}\rangle$ is a linear combination of the bulk microstates inside the polyhedron $\fp$.

The spatial region $\Sig$ has been discretized by a large number of polyhedra $\fp$, the semiclassical geometry of $\Sig$ endows the polyhedron geometry to each $\cp$. In particular, the semiclassical geometry endows the area $\Ar_f$ to each polyhedron face $f$. Thus in order that $\Phi_b$ represents the semiclassical geometry, the following constraint has to be imposed to the microstates in the linear combination: recall that $|\mu_f\rangle =\otimes_\ell |j_l,m_l\rangle $, thus the sum of all quantum areas on edges $\ell$ intersecting $f$ is constrained to be the total area $\Ar_f$, being the area of polyhedron face $f$ endowed by the semiclassical geometry
\be
8\pi\g\ell_P^2\sum^{N_\G(f)}_{l,l\cap f\neq\emptyset}\sqrt{j_l(j_l+1)}=\Ar_f.\label{coarse}
\ee

The above constraint fixes the total number of states $|\mu_f\rangle$ at $f$, which is the range of the sum over $\mu_f$ in $\Phi_b$ Eq.\Ref{cla}. $D_f$ is the number of microstates $|\mu_f\rangle$ satisfying the constraint Eq.\Ref{coarse}
\be
D_f=\sum_{N_\G(f)}\sum_{\{j_l\}}^{\Ar_f}\prod^{N_\G(f)}_l(2j_l+1).\label{Df}
\ee 

For each polyhedron $\fp$ , one imposes the constraint Eq.\Ref{coarse} to the microstate $|\mu_f\rangle\in\ch_\partial(f)$ at the polyhedron faces. In principle the polyhedron geometry of $\fp$ endowed by bulk semiclassical geometry also constrains the bulk microstates $|\xi_\fp\rangle$ inside $\fp$, since these states relate to the shape and curvatures of $\fp$. However, it turns out in Section \ref{EE} that in $\Phi_b$, the detailed knowledge of the geometrical constraint on $|\xi_\fp\rangle$ doesn't affect the computation that recovers the Ryu-Takayanagi formula of holographic entanglement entropy. So we only explicitly impose Eq.\Ref{coarse} to $\Phi_b$ in Eq.\Ref{cla}, while keeping the other geometrical constraints implicit. The following derivation works for all possible $\lt|\phi(\fp)_{\vec{\mu}}\rt\rangle$ in the $\Phi_b$, subject to Eq.\Ref{coarse}.

\section{Bond Dimension}\label{bond}

In this section, we estimate $D_f$ the dimension of $\ch_\partial(f)$ at each face $f$. The following estimation is an important step in deriving the Ryu-Takayanagi formula of holographic entanglement entropy. 

The counting of states Eq.\Ref{Df} has been studied in the literature of LQG black hole entropy counting. See e.g. \cite{GP2011,Ghosh:2004wq,QGandBH}. It is interesting to see that the boundary degree of freedom of the coarse grained spin-networks relates to the horizon degree of freedom of a quantum black hole. As we will see in the following, it provides nontrivial implication to the entanglement entropy of the boundary state at $\partial\Sig$. Following \cite{GP2011,Ghosh:2004wq} we define $n_j$ to be the number of intersecting edges $l$ carrying spin $j$. The dimension $D_f$ can then be written as
\be
D_f=\sum_{N_\G(f)}\sum_{\{n_j\}} d[\{n_j\}],\quad d[\{n_j\}]=\lt(\sum_j n_j !\rt)\prod_j\frac{(2j+1)^{n_j}}{n_j!}
\ee
where $\sum_{\{n_j\}}$ is subjected to the constraint that
\be
C_1=\sum_{j}\sqrt{j(j+1)}\, n_j-\frac{\Ar_{f}}{8\pi\g\ell_P^2}=0,\quad C_2=\sum_j n_j-N_{\G}(f)=0.\label{condition}
\ee

The main contribution of $D_f$ clearly comes from the regime that $N_\G(f)\gg1$. We assume $N_\G(f)_{\text{max}}=\frac{\Ar_f}{4\pi\g\ell_P^2\sqrt{3}}\gg 1$. For fixed $N_\G(f)$, the constrained sum $\sum_{\{n_j\}} d[\{n_j\}]$ is equivalent to the statistical ensemble of $\cn=N_\G(f)$ identical systems with total energy $\ce=\frac{\Ar_{f}}{8\pi\g\ell_P^2}$. Here we have assumed that all intersections of $l\cap f$ are distinguishable. 

It is standard that as $N_{\G}(f)\gg1$, the number of micro-states $\sum_{\{n_j\}} d[\{n_j\}]$ in the ensemble is dominated by the contribution from the configuration $\{\bar{n}_j\}$ which maximizes $d[\{{n}_j\}]$ \cite{pathria}. $\{\bar{n}_j\}$ is the solution of the variational equation $\delta\ln d[\{n_j\}]-\b \delta C_1-\mu\delta C_2=0$ where $\b,\mu$ are two Lagrangian multipliers. Under Stirling's approximation, we obtain
\be
\frac{\bar{n}_j}{N_{\G}(f)}=(2j+1)e^{-\b\sqrt{j(j+1)}-\mu}. 
\ee
The Lagrangian multipliers $\b,\mu$ are then determined by plugging the solution into the constraints
\be
\sum_{j} (2j+1)e^{-\b\sqrt{j(j+1)}-\mu}\sqrt{j(j+1)}=\frac{\Ar_{f}}{8\pi\g\ell_P^2 N_{\G}(f)},\quad 
\sum_j (2j+1)e^{-\b\sqrt{j(j+1)}}=e^{\mu}.
\ee
As a result, in the regime $N_{\G}(f)\gg1$, 
\be
D_f\simeq \sum_{N_\G(f)}\exp\lt[\frac{\b}{8\pi\g\ell_P^2}\,\Ar_{f}+ N_{\G}(f)\,\mu(\b)\rt]
\ee
Here $\mu(\b)=\ln\lt[\sum_j (2j+1)e^{-\b\sqrt{j(j+1)}}\rt]$ is a statistical averaging of the dimension of SU(2) spin-$j$ irrep. The above result is an analog of the entropy formula in the ensemble with $\cn$ identical systems and total energy $\ce$: $S(\ce,\cn)=\b \ce+\mu \cn$. We see that by viewing the above counting of states as a statistical canonical ensemble of $N_{\G}(f)$ punctures on the surfaces, we may understand $\b$ as an effective temperature and $\mu$ as an effective chemical potential. By the variation $\delta S(\ce,\cn)=\b \delta\ce+\mu \delta\cn$ ($\delta S/\delta\b=0$), we see that $S(\ce,\cn)$ is monotonically increasing as $\cn$ increases, when $\mu>0$. The maximum happens at $\delta S/\delta \cn=\mu(\b)=0$, where $\delta^2S/\delta \cn^2\sim -1/\cn$
\footnote{$\frac{\delta^2S}{\delta \cn^2}=\frac{\delta\mu}{\delta\cn}=\frac{\delta\mu}{\delta\b}\frac{\delta\b}{\delta\cn}$. At $\b=\b_0$ where $\mu(\b_0)=0$, $\frac{\delta\mu}{\delta\b}=-\frac{\sum_j  (2j+1)e^{-\b\sqrt{j(j+1)}}\sqrt{j(j+1)}}{\sum_j (2j+1)e^{-\b\sqrt{j(j+1)}}}=-\frac{\ce}{\cn}$, and $-\delta\b\sum_{j} (2j+1)e^{-\b\sqrt{j(j+1)}}{j(j+1)}=-\frac{\ce}{\cn^2}\delta\cn$, i.e. $\frac{\delta\b}{\delta\cn}=\frac{\ce}{\lag\ce^2\rag}$. Thus $\frac{\delta^2S}{\delta \cn^2}=-\frac{1}{\cn}\frac{\ce^2}{\lag\ce^2\rag}$. Expanding $S$ up to the quadratic order, the sum can be approximated by a Gauss integral
\be
D_f\simeq e^{\b_0\ce}\sum_{\cn}\exp\lt[{-\cn\frac{\ce^2}{\lag\ce^2\rag}\lt(\frac{\delta\cn}{\cn}\rt)^2}\rt]\simeq e^{\b_0\ce+\cdots}.
\ee 
where $\cdots$ stands for corrections of order $\ln\cn\leq \ln\lt(\frac{\Ar_{f}}{4\pi\g\ell_P^2\sqrt{3}}\rt) $.}. 
Therefore we obtain that the bond dimension $D_f$ behaves as the exponential of the face area.
\be
D_f\simeq \exp\lt[\frac{\b_0}{8\pi\g\ell_P^2}\,\Ar_{f}+\cdots\rt]\label{arealaw}
\ee
where $\cdots$ stands for the logarithmic corrections in $\ln\lt(\frac{\Ar_{f}}{8\pi\g\ell_P^2}\rt)$. $\b_0$ is a universal constant (independent of $f$), being the solution to 
\be
\mu(\b_0)=\ln\lt[\sum_j (2j+1)e^{-\b_0\sqrt{j(j+1)}}\rt]=0. \quad 2\pi\b_0\simeq 0.274...
\ee
The prefactor $\frac{\b_0}{8\pi\g\ell_P^2}$ has been suggested to be identified as the IR value of $1/4 G_N$ \cite{QGandBH,Ghosh:2012wq}.

\section{Holographic Entanglement Entropy and Random Tensor}\label{EE}

As it is shown in Section \ref{EHM}, the LQG states with boundary is a concrete realization of exact holographic mapping in \cite{Qi2,Qi1}. Therefore the similar techniques in \cite{Qi1} can be imported here to study the holographic entanglement entropy of boundary state.

One of the key technique in \cite{Qi1}, which is employed here, is to take $|\cv_\fp\rangle$ at each polyhedron $\fp$ as a random state in $\ch_\fp$. We may define an arbitrary reference state $|0_\fp\rangle\in\ch_\fp$, so that $|\cv_\fp\rangle= U|0_\fp\rangle$ with $U$ a unitary operator. For any function $f(|\cv_\fp\rangle)$, e.g. the entanglement entropy computed from $|\cv_\fp\rangle$, the state $|\cv_\fp\rangle$ being random implies that $f(|\cv_\fp\rangle)$ should be random averaged according to the Haar probability measure $\rmd U$. The Haar probability measure is normalized $\int\rmd U=1$ and invariant under unitary transformation. Therefore we always consider the random averaged quantity 
\be
\overline{f(|\cv_\fp\rangle)}:= \int\rmd U\, f(|\cv_\fp\rangle)=\int\rmd U\, f\lt( U|0_\fp\rangle\rt).
\ee
This random averaging may be viewed as a part of coarse graining procedure, which smears the microscopic details (of Planck scale) within each semiclassical polyhedron $\fp$. The above random state technique has a long and rich history in quantum information theory (see e.g \cite{randomrev} for a review), and has been often used in the studies of entanglement entropy (see e.g. \cite{hayden,Page:1993df}).   

Although our derivation of holographic entanglement entropy follows the similar routine as in \cite{Qi1}, there are some key difference and improvement, coming from the feature of LQG.
\begin{itemize}

\item Thanks to LQG and its interpretation as quantum geometry, the bond dimension $D_f$ relates to the face area $\Ar_f$ of polyhedron by Eq.\Ref{arealaw}. It is one of the key ingredients to relate the boundary entanglement entropy to a path integral of Nambu-Goto action $S_{NG}\propto \Ar_\cs/\ell_P^2$ of bulk 2-surface $\cs$. The Ryu-Takayanagi surface with minimal area stands out as the minimum of the Nambu-Goto action. The leading contribution of the path integral in the semiclassical limit $\ell_P\to 0$ reproduces precisely the Ryu-Takayanagi formula of holographic entanglement entropy.

\item In the technical aspect, the Hilbert space $\ch_\fp$ from LQG is of different structure to the vertex Hilbert space $\ch_x$ (where $|V_x\rangle$ leaves) proposed in \cite{Qi1}. $\ch_x$ in \cite{Qi1} has been assumed to be a tensor product of the bulk and boundary Hilbert spaces. However in our context, $\ch_\fp$ defined in Eq.\Ref{chp} doesn't have a pure tensor product structure, i.e. $\ch_\fp$ cannot be written as a tensor product between $\ch_b(\fp)$ and $\ch_\partial (f)$. The reason is that for LQG states, the bulk intertwiners $I_v$ close to the boundary depends on the boundary spins $j_l$. Thus the states $|\cv_\fp\rangle\in\ch_\fp$ always exhibit certain entanglement between $\ch_b(\fp)$ and $\ch_\partial (f)$ (see Eq.\Ref{coarsecv}).  $\ch_\fp$ where the random state $|\cv_\fp\rangle$ lives, is the domain where the random averaging is carried out. Therefore the difference with $\ch_x$ in \cite{Qi1} results in some technical differences to \cite{Qi1} in the random averaging procedure. We will come back to this point in a moment.

\end{itemize}

\subsection{Second Renyi Entropy}\label{2nd}

Recall the boundary state $\Psi$ obtained from exact holographic mapping Eq.\Ref{PhiEV}. We define the density matrix of $\Psi$ by 
\be
\rho=|\Psi\rangle\langle\Psi|&=&\tr_{\ch_b\otimes\ch_E}\lt(\rho_P\otimes_\fp|\cv_\fp\rangle\langle \cv_\fp|\rt)=\langle \Phi_b|\langle E|\Big(\otimes_\fp|\cv_\fp\rangle\langle \cv_\fp|\Big) |\Phi_b\rangle|E\rangle
\ee
where $\rho_P=|\Phi_b\rangle\langle \Phi_b|\otimes|E\rangle\langle E|$.

We divide $\partial\Sig$ into two regions $A$ and $\bA$, and define the reduced density matrix $\rho_A$ by tracing out the states in $\bA$. Here the region A is always set to be composed of a (large) multiple of polyhedron faces. We set the regions $A$ and $\bA$ in such a way that there is no polyhedron adjacent to $\partial\Sig$, containing faces both in $A$ and $\bA$. In other words, $A$ and $\bA$ connect to two different sets of polyhedra with no common element.

In this subsection we study the second Renyi entropy $S_2(A)=-\ln\lt[\tr\rho_A^2/(\tr\rho_A)^2\rt]$, and its random average $\overline{S_2(A)}$. It is shown in \cite{Qi1} that the averaged second Renyi entropy can be approximated with high prevision to leading order in the large bond dimension limit, by the separate averages $\overline{\tr\rho_A^2}$ and $\overline{(\tr\rho_A)^2}$
\be
\overline{S_2(A)}\simeq -\ln\frac{\overline{\tr\rho_A^2}}{\overline{(\tr\rho_A)^2}}.
\ee
The error is suppressed when the bond dimension $D_f$ is large, which is indeed true by Eq.\Ref{arealaw} in the semiclassical regime $\Ar_f\gg\ell_P^2$. The above relation can be generalized to the Renyi entropy of arbitrary order, namely
\be
\overline{S_n(A)}\simeq -\frac{1}{n-1}\ln\frac{\overline{\tr\rho_A^n}}{\overline{(\tr\rho_A)^n}}
\ee

Firstly we compute $\overline{\tr\rho_A^2}$. Recall that a basis in $\ch_{\partial}$ may be chosen as $|\mu_f\rangle_A|\mu_f\rangle_\bA\equiv \otimes_{f\subset A}|\mu_f\rangle_A\otimes_{f\subset \bA}|\mu_f\rangle_\bA$, where $\mu_f=1\cdots \dim\ch_\partial(f)$, we may write\footnote{This relation can be checked straight-forwardly using the definition of $\cf_A$:
\be
\tr_{\ch_{\partial}\otimes\ch_\partial}\lt[(\rho\otimes\rho)\cf_A\rt]
&=&\Big(\langle \mu_f|_A\langle \mu_f|_\bA\Big)\otimes\lt(\langle \mu'_f|_A\langle \mu'_f|_\bA\rt)\, \Big[(\rho\otimes\rho)\cf_A\Big]\,\Big(|\mu_f\rangle_A|\mu_f\rangle_\bA\Big)\otimes\lt(|\mu'_f\rangle_A|\mu_f'\rangle_\bA\rt)\nonumber\\
&=&\Big(\langle \mu_f|_A\langle \mu_f|_\bA\Big)\otimes\lt(\langle \mu'_f|_A\langle \mu'_f|_\bA\rt)\,\Big[\rho\otimes\rho\Big]\,\Big(|\mu'_f\rangle_A|\mu_f\rangle_\bA\Big)\otimes\lt(|\mu_f\rangle_A|\mu'_f\rangle_\bA\rt)\nonumber\\
&=&\Big(\langle \mu_f|_A\langle \mu_f|_\bA\Big)\,\rho\,\Big(|\mu'_f\rangle_A|\mu_f\rangle_\bA\Big)\ 
\lt(\langle \mu'_f|_A\langle \mu'_f|_\bA\rt)\,\rho\,\lt(|\mu_f\rangle_A|\mu'_f\rangle_\bA\rt)\nonumber\\
&=&\langle \mu_f|\rho_A|\mu'_f\rangle_A\ 
\langle \mu'_f|\rho_A|\mu_f\rangle_A=\tr_A\rho_A^2
\ee} 
\be
\tr\rho_A^2=\langle \mu_f|\rho_A|\mu'_f\rangle_A\ \langle \mu'_f|\rho_A|\mu_f\rangle_A=\tr_{\ch_{\partial}\otimes\ch_\partial}\lt[(\rho\otimes\rho)\cf_A\rt],\label{rho2}
\ee
where repeating labels means summing over the labels. $\cf_A$ is a swapping operator acting on $\ch_\partial\otimes\ch_\partial$, swapping the states in $A$:
\be
\cf_A\Big(|\mu_f\rangle_A|\mu_f\rangle_\bA\Big)\otimes\lt(|\mu'_f\rangle_A|\mu'_f\rangle_\bA\rt)=\Big(|\mu'_f\rangle_A|\mu_f\rangle_\bA\Big)\otimes\lt(|\mu_f\rangle_A|\mu'_f\rangle_\bA\rt)
\ee
Using the definition of $\rho$, 
\be
\tr\rho_A^2&=&\tr_{\ch_{\partial}\otimes\ch_\partial}\ \tr_{\ch_b\otimes\ch_E}\otimes\tr_{\ch_b\otimes\ch_E}\lt[\Big(\lt(\rho_P\otimes_\fp|\cv_\fp\rangle\langle \cv_\fp|\rt)\otimes\lt(\rho_P\otimes_\fp|\cv_\fp\rangle\langle \cv_\fp|\rt)\Big)\cf_A\rt].
\ee

We consider $|\cv_\fp\rangle$ as a random state in $\ch_\fp$ at each $\cp$. The random average of $|\cv_\fp\rangle$ gives the following simple result by Schur's Lemma \cite{Qi1,church}
\be
\overline{|\cv_\fp\rangle\langle \cv_\fp|\otimes |\cv_\fp\rangle\langle \cv_\fp|}&=&\int\rmd U\, (U\otimes U)\,|0_\fp\rangle\langle 0_\fp|\otimes |0_\fp\rangle\langle 0_\fp|\,(U^\dagger\otimes U^\dagger)\nonumber\\
&=&\frac{I_\fp+\cf_\fp}{\dim(\ch_\fp)^2+\dim(\ch_\fp)}\in\ch_\fp\otimes\ch_\fp\otimes\ch_\fp^*\otimes\ch_\fp^*.\label{IF}
\ee 
The action of $I_\fp$ and $\cf_\fp$ are the identity and swapping operators
\be
I_\fp\,| \xi^{\vec{\mu}}_\fp,\mu_f\rangle\otimes | \xi'^{\vec{\mu}'}_\fp,\mu'_f\rangle&=&|  \xi_\fp^{\vec{\mu}},\mu_f\rangle\otimes |  \xi_\fp'^{\vec{\mu}'},\mu'_f\rangle\nonumber\\
\cf_\fp\,| \xi_\fp^{\vec{\mu}},\mu_f\rangle\otimes |  \xi_\fp'^{\vec{\mu}'},\mu'_f\rangle&=& | \xi_\fp'^{\vec{\mu}'},\mu'_f\rangle\otimes| \xi_\fp^{\vec{\mu}},\mu_f\rangle.
\ee
Note that here the Hilbert space $\ch_\fp$ is in principle infinite-dimensional. But we made a regularization by cutting-off the dimension to be finite. It turns out that the resulting Renyi entropy is independent of the dimension of $\ch_\fp$, so we can freely remove the cut-off. 

From the above action of $I_\fp$ and $\cf_\fp$, one may see the difference with \cite{Qi1}. Here the Hilbert space $\ch_\fp$ cannot be factorized as the tensor product of $\ch_b(\fp)$ and $\ch_\partial(f)$. In general the states acted by $I_\fp$ and $\cf_\fp$ are not pure tensor product between $| \xi^{\vec{\mu}}_\fp\rangle$ and $|\mu_f'\rangle$, but rather the entangled states $| \xi^{\vec{\mu}}_\fp,\mu_f\rangle$ with the correlation of $\vec{\mu}$. Therefore the operators $I_\fp$ and $\cf_\fp$ cannot be factorized into the identities and swappings in the individual $\ch_b(\fp)$ and $\ch_\partial(f)$, which is different from the situation assumed in \cite{Qi1}.

By the random average Eq.\Ref{IF}, the average $\overline{\tr\rho_A^2}$ becomes a sum of $2^{(\#\ \text{of}\ \fp)}$ terms. Each term corresponds to a choice of $I_\fp$ or $\cf_\fp$ at each $\fp$. As in \cite{Qi1}, we introduce an Ising variable $s_\fp=1$ (or $s_\fp=-1$) to label the choice of $I_\fp$ (or $\cf_\fp$) at $\fp$. Given a choice $\{s_\fp\}$ at each term of $\overline{\tr\rho_A^2}$, the corresponding term reads
\be
\langle \mu_f|_A\langle \mu_f|_\bA\langle\Phi_b|\langle E|\otimes\langle \mu'_f|_A\langle \mu'_f|_\bA\langle \Phi_b|\langle E|\prod_{\fp, s_\fp=1}I_\fp\prod_{\fp, s_\fp=-1}\cf_\fp\ |\Phi_b\rangle|E\rangle|\mu'_f\rangle_A|\mu_f\rangle_\bA\otimes |\Phi_b\rangle|E\rangle|\mu_f\rangle_A|\mu'_f\rangle_\bA.\label{sandwich}
\ee

Here a generic bulk state $|\Phi_b\rangle\in\ch_b$ can be written as
\be
|\Phi_b\rangle=\sum_{\vec{\mu}}|\Phi_{\vec{\mu}}\rangle=\sum_{\xi_\fp,\vec{\mu}}\Phi_{\vec{\xi},\vec{\mu}}\otimes_\fp|\xi_\fp^{\vec{\mu}}\rangle.\label{genPhib}
\ee 
We assume $|\Phi_b\rangle$ satisfy Eq.\Ref{cla} with factorized coefficients $\Phi_{\vec{\xi},\vec{\mu}}$ 
\be
|\Phi_b\rangle=\sum_{\{\mu_f\}}^{\{D_f\}}|\Phi_{\vec{\mu}}\rangle=\sum_{\{\mu_f\}}^{\{D_f\}}\sum_{\xi_\fp}\Phi_{\vec{\xi},\vec{\mu}}\otimes_\fp|\xi_\fp^{\vec{\mu}}\rangle,\quad \Phi_{\vec{\xi},\vec{\mu}}=\prod_\fp\phi(\fp)_{\xi_\fp,\vec{\mu}}.\label{factor}
\ee
As it is discussed above, $\Phi_b$ of the above type represents the locality. Its image under holographic mapping is a tensor network state as a representation of boundary CFT ground state. $|\phi(\fp)_{\vec{\mu}}\rangle=\sum_{\xi_\fp}\phi(\fp)_{\xi_\fp,\vec{\mu}}|\xi_\fp^{\vec{\mu}}\rangle$ is assumed to be normalized for any $\vec{\mu}$:
\be
\sum_{\xi_\fp}\phi(\fp)_{\xi_\fp,\vec{\mu}}^*\phi(\fp)_{\xi_\fp,\vec{\mu}}=1.\label{normalization}
\ee
$D_f$ gives the bond dimension of the resulting tensor network $\Psi$ as the image of the exact holographic mapping.

We compute the operator $\prod_{\fp, s_\fp=1}I_\fp\prod_{\fp,\ s_\fp=-1}\cf_\fp$ acting on the right in Eq.\Ref{sandwich}: 
\be
&&\prod_{\fp, s_\fp=1}I_\fp\prod_{\fp,\ s_\fp=-1}\cf_\fp\ |\Phi_b\rangle|E\rangle|\mu'_f\rangle_A|\mu_f\rangle_\bA\otimes |\Phi_b\rangle|E\rangle|\mu_f\rangle_A|\mu'_f\rangle_\bA\nonumber\\
&=&\prod_{\fp, s_\fp=1}I_\fp\prod_{\fp,\ s_\fp=-1}\cf_\fp \sum_{{\xi}_\fp,\mu_f}\Phi_{\vec{\xi},\mu_f,\{\mu_f'\}_A,\{\mu_f\}_\bA}\otimes_\fp\lt|\xi_\fp^{\vec{\mu}},\mu_f,\{\mu_f'\}_A,\{\mu_f\}_\bA\rag\otimes\sum_{{\xi}'_\fp,\mu_f'}\Phi_{\vec{\xi}',{\mu}_f',\{\mu_f\}_A,\{\mu'_f\}_\bA}\otimes_\fp\lt|\xi_{\fp}'^{\vec{\mu}'},\mu'_f,\{\mu_f\}_A,\{\mu'_f\}_\bA\rag\nonumber\\
&=&\sum_{{\xi}_\fp,\mu_f}\sum_{{\xi}'_\fp,\mu_f'}\Phi_{\vec{\xi},{\mu}_f,\{\mu_f'\}_A,\{\mu_f\}_\bA}\Phi_{\vec{\xi}',{\mu}'_f,\{\mu_f\}_A,\{\mu'_f\}_\bA}
\prod_{\fp,s_\fp=1}\lt|\xi_\fp^{\vec{\mu}},\mu_f,\{\mu_f'\}_A,\{\mu_f\}_\bA\rag\otimes\lt|\xi_{\fp}'^{\vec{\mu}'},\mu'_f,\{\mu_f\}_A,\{\mu'_f\}_\bA\rag\nonumber\\
&&\quad\prod_{\fp, s_\fp=-1}\lt|\xi_{\fp}'^{\vec{\mu}'},\mu'_f,\{\mu_f\}_A,\{\mu'_f\}_\bA\rag\otimes\lt|\xi_\fp^{\vec{\mu}},\mu_f,\{\mu_f'\}_A,\{\mu_f\}_\bA\rag
\ee
In the first step, we use the fact that $I_\fp,\cf_\fp$ only act on the states $| \xi_\fp^{\vec{\mu}}\rangle\otimes|\mu_f\rangle$ where the $\vec{\mu}$ labels of $\xi_\fp^{\vec{\mu}}$ coincide with the $\mu_f$ labels in $|\mu_f\rangle$, while projecting out the states which doesn't satisfy this coincidence. 

We take the inner product and compute the term \Ref{sandwich} in $\overline{\tr\rho_A^2}$
\be
&&\sum_{{\zeta}_\fp,\nu_f}\sum_{{\zeta}'_\fp,\nu_f'}\sum_{{\xi}_\fp,\mu_f}\sum_{{\xi}'_\fp,\mu_f'}\sum_{\{\mu_f\}_A}\sum_{\{\mu'_f\}_A}\sum_{\{\mu_f\}_\bA}\sum_{\{\mu'_f\}_\bA}\Phi_{\vec{\zeta},{\nu}_f,\{\mu_f\}_A,\{\mu_f\}_\bA}^*\Phi_{\vec{\zeta}',{\nu}'_f,\{\mu_f'\}_A,\{\mu_f'\}_\bA}^*\Phi_{\vec{\xi},{\mu}_f,\{\mu_f'\}_A,\{\mu_f\}_\bA}\Phi_{\vec{\xi}',{\mu}'_f,\{\mu_f\}_A,\{\mu'_f\}_\bA}\nonumber\\
&&\prod_{\fp,s_\fp=1}\lag\zeta_\fp,\nu_f,\{\mu_f\}_A,\{\mu_f\}_\bA
\Big|\xi_\fp,\mu_f,\{\mu'_f\}_A,\{\mu_f\}_\bA\rag
\lag\zeta'_\fp,\nu'_f,\{\mu_f'\}_A,\{\mu_f'\}_\bA
\Big|\xi'_\fp,\mu'_f,\{\mu_f\}_A,\{\mu'_f\}_\bA\rag\nonumber\\
&&\prod_{\fp, s_\fp=-1}\lag\zeta_\fp,\nu_f,\{\mu_f\}_A,\{\mu_f\}_\bA\Big|\xi'_\fp,\mu'_f,\{\mu_f\}_A,\{\mu'_f\}_\bA\rag
\lag\zeta'_\fp,\nu'_f,\{\mu'_f\}_A,\{\mu'_f\}_\bA\Big|\xi_\fp,\mu_f,\{\mu'_f\}_A,\{\mu_f\}_\bA\rag\nonumber\\
&=&\sum_{{\zeta}_\fp,\nu_f}\sum_{{\zeta}'_\fp,\nu_f'}\sum_{{\xi}_\fp,\mu_f}\sum_{{\xi}'_\fp,\mu_f'}\sum_{\{\mu_f\}_A}\sum_{\{\mu'_f\}_A}\sum_{\{\mu_f\}_\bA}\sum_{\{\mu'_f\}_\bA}\Phi_{\vec{\zeta},{\nu}_f,\{\mu_f\}_A,\{\mu_f\}_\bA}^*\Phi_{\vec{\zeta}',{\nu}'_f,\{\mu_f'\}_A,\{\mu_f'\}_\bA}^*\Phi_{\vec{\xi},{\mu}_f,\{\mu_f'\}_A,\{\mu_f\}_\bA}\Phi_{\vec{\xi}',{\mu}'_f,\{\mu_f\}_A,\{\mu'_f\}_\bA}\nonumber\\
&&\prod_{\fp,s_\fp=1}\delta_{\zeta_\fp,\xi_\fp}\delta_{\zeta_\fp',\xi_\fp'}\delta_{\nu_f\mu_f}\delta_{\nu_f'\mu_f'}\delta_{\{\mu_f\}_A,\{\mu'_f\}_A}\delta_{\{\mu'_f\}_A,\{\mu_f\}_A}
\prod_{\fp, s_\fp=-1}\delta_{\zeta_\fp,\xi_\fp'}\delta_{\zeta_\fp',\xi_\fp}\delta_{\nu_f\mu_f'}\delta_{\nu_f'\mu_f}\delta_{\{\mu_f\}_\bA,\{\mu_f'\}_\bA}\delta_{\{\mu_f'\}_\bA,\{\mu_f\}_\bA}.\label{term}
\ee

We firstly fix the $\mu_f,\nu_f,\mu_f',\nu_f'$ labels, and carry out the sum over $\xi_\fp,\xi_\fp',\zeta_\fp,\zeta_\fp'$:
\be
&&\sum_{{\zeta}_\fp,{\zeta}'_\fp,{\xi}_\fp,{\xi}'_\fp}\Phi_{\vec{\zeta},{\nu}_f,\{\mu_f\}_A,\{\mu_f\}_\bA}^*\Phi_{\vec{\zeta}',{\nu}'_f,\{\mu_f'\}_A,\{\mu_f'\}_\bA}^*\Phi_{\vec{\xi},{\mu}_f,\{\mu_f'\}_A,\{\mu_f\}_\bA}\Phi_{\vec{\xi}',{\mu}'_f,\{\mu_f\}_A,\{\mu'_f\}_\bA}\prod_{\fp,s_\fp=1}\delta_{\zeta_\fp,\xi_\fp}\delta_{\zeta_\fp',\xi_\fp'}
\prod_{\fp, s_\fp=-1}\delta_{\zeta_\fp,\xi_\fp'}\delta_{\zeta_\fp',\xi_\fp}\nonumber\\
&=&\exp\lt[-S_2\lt(\{s_\fp=-1\};\Phi_{\vec{\mu}}\rt)\rt].
\ee
where $S_2\lt(\{s_\fp=-1\};\Phi_{\vec{\mu}}\rt)$ is the second Renyi entropy in the $s_\fp=-1$ domain for the state $|\Phi_{\vec{\mu}}\rangle$ in Eq.\Ref{genPhib} with fixed $|\mu_f\rangle$ at each $f$. If $\Phi_b$ satisfies Eq.\Ref{factor}, $\{D_f\}$ give a cut-offs for the sums over $\mu_f,\mu'_f,\nu_f,\nu'_f$.  We also have $S_2\lt(\{s_\fp=-1\};\Phi_{\vec{\mu}}\rt)=0$:
\be
&&\sum_{{\zeta}_\fp,{\zeta}'_\fp,{\xi}_\fp,{\xi}'_\fp}\prod_{\fp}\phi(\fp)_{{\zeta}_\fp,{\nu}_f,\{\mu_f\}_A,\{\mu_f\}_\bA}^*\phi(\fp)_{{\zeta}'_\fp,{\nu}'_f,\{\mu_f'\}_A,\{\mu_f'\}_\bA}^*\phi(\fp)_{{\xi}_\fp,{\mu}_f,\{\mu_f'\}_A,\{\mu_f\}_\bA}\phi(\fp)_{{\xi}'_\fp,{\mu}'_f,\{\mu_f\}_A,\{\mu'_f\}_\bA}\prod_{\fp,s_\fp=1}\delta_{\zeta_\fp,\xi_\fp}\delta_{\zeta_\fp',\xi_\fp'}
\prod_{\fp, s_\fp=-1}\delta_{\zeta_\fp,\xi_\fp'}\delta_{\zeta_\fp',\xi_\fp}\nonumber\\
&=&\sum_{{\zeta}_\fp,{\zeta}'_\fp,{\xi}_\fp,{\xi}'_\fp}\prod_{\fp,s_\fp=1}\phi(\fp)_{{\zeta}_\fp,{\mu}_f,\{\mu_f\}_A,\{\mu_f\}_\bA}^*\phi(\fp)_{{\zeta}'_\fp,{\mu}'_f,\{\mu_f\}_A,\{\mu_f'\}_\bA}^*\phi(\fp)_{{\xi}_\fp,{\mu}_f,\{\mu_f\}_A,\{\mu_f\}_\bA}\phi(\fp)_{{\xi}'_\fp,{\mu}'_f,\{\mu_f\}_A,\{\mu'_f\}_\bA}\delta_{\zeta_\fp,\xi_\fp}\delta_{\zeta_\fp',\xi_\fp'}\nonumber\\
&&\prod_{\fp, s_\fp=-1}\phi(\fp)_{{\zeta}_\fp,{\mu'}_f,\{\mu_f\}_A,\{\mu_f\}_\bA}^*\phi(\fp)_{{\zeta}'_\fp,{\mu}_f,\{\mu_f'\}_A,\{\mu_f\}_\bA}^*\phi(\fp)_{{\xi}_\fp,{\mu}_f,\{\mu_f'\}_A,\{\mu_f\}_\bA}\phi(\fp)_{{\xi}'_\fp,{\mu}'_f,\{\mu_f\}_A,\{\mu_f\}_\bA}\delta_{\zeta_\fp,\xi_\fp'}\delta_{\zeta_\fp',\xi_\fp}\nonumber\\
&=&1
\ee
In the first step, we take advantage of the delta's of $\mu_f,\nu_f$ labels in Eq.\Ref{term}, and replaces the labels. In the second step, we use the normalization Eq.\Ref{normalization}. In the following we focus on this type of bulk state $\Phi_b$, with vanishing bulk entanglement entropy. 


We denote the $s_\fp=1$ ($s_\fp=-1$) region by $R_+$ ($R_-$) as a close subdomain of $\Sig$. The term in Eq.\Ref{term} reduces to the following contribution
\be
&&
\sum_{\nu_f}\sum_{\nu_f'}\sum_{\mu_f}\sum_{\mu_f'}\sum_{\{\mu_f\}_A}\sum_{\{\mu'_f\}_A}\sum_{\{\mu_f\}_\bA}\sum_{\{\mu'_f\}_\bA}\prod_{\fp,s_\fp=1}\delta_{\nu_f\mu_f}\delta_{\nu_f'\mu_f'}\delta_{\{\mu_f\}_A,\{\mu'_f\}_A}\delta_{\{\mu'_f\}_A,\{\mu_f\}_A}
\prod_{\fp, s_\fp=-1}\delta_{\nu_f\mu_f'}\delta_{\nu_f'\mu_f}\delta_{\{\mu_f\}_\bA,\{\mu_f'\}_\bA}\delta_{\{\mu_f'\}_\bA,\{\mu_f\}_\bA}\nonumber\\
&=&
\prod_{f\subset R_+\setminus \partial R_+}D_f^2\prod_{f\subset R_-\setminus \partial R_-}D_f^2\prod_{f\subset R_+\cap R_-} D_f\prod_{f\subset R_+\cap A}D_f\prod_{f\subset R_+\cap \bA}D_f^2\prod_{f\subset R_-\cap A}D_f^2\prod_{f\subset R_-\cap \bA}D_f,
\ee
because the cut-offs for the sums over $\mu_f,\mu'_f,\nu_f,\nu'_f$ have been introduced by $\Phi_b$. One may write the above result in a form as $e^{-A[s_\fp]}$ whose ``effective action'' $A[s_\fp]$ reads
\be
A[s_\fp]=-\sum_{f\ \text{bulk}}\frac{1}{2}\ln D_f(s_\fp s_{\fp'}-1)-\sum_{f\ \text{boundary}}\half\ln D_f(h_\fp s_\fp-1)
+\text{const.}
\ee
The bulk $f$ is the interface between $\fp,\fp'$ in the first term, while the boundary $f$ is a face of $\fp$ in the second term. $h_\fp$ is a ``boundary field'' satisfying $h_\fp=1$ $(h_\fp=-1)$ as $\fp$ close to $\bA$ ($\fp$ close to $A$). The non-explicit constant terms doesn't depends on the Ising variables $s_\fp$, but depends on $D_f$. By summing of the terms over all possible Ising configurations $\{s_\fp\}$, it reproduces the Ising model as in \cite{Qi1}. But now it comes with the non-uniform Ising couplings $\ln D_f$ interpreted as face areas $\Ar_f/\ell_P^2$ by Eq.\Ref{arealaw}.


The semiclassical regime $\Ar_f\gg\ell_P^2$ implies $D_f\gg1$. For a given region $A\subset \partial\Sig$, when we write $\overline{\tr\rho_A^2}$ as a sum over Ising configurations $\{s_\fp\}$, the sum is dominated by the Ising configurations $\{\bar{s}_\fp\}$ such that $R_+\cap A=\emptyset$ and $R_-\cap \bA=\emptyset$\footnote{The domain-wall Ising configurations are selected by maximizing the number of $D_f^2$ factors. It may also be understood from the Ising action $A[s_\fp]$, in which the dominant configuration should minimize $A[s_\fp]$.}. In other words, $R_+$ ($R_-$) is bounded by $\bA$ ($A$). The interface (domain-wall) $\cs=R_+\cap R_-$ is attached at the boundary of $A$, i.e. $\partial \cs=\partial A$.
\be
\overline{\tr\rho_A^2}\simeq\prod_\fp\frac{1}{\dim(\ch_\fp)^2+\dim(\ch_\fp)}\sum_{\{\bar{s}_\fp\}}
\prod_{f\not\subset R_+\cap R_-}D_f^2\prod_{f\subset R_+\cap R_-} D_f.
\ee

$\overline{(\tr\rho_A)^2}$ can be computed in the similar way as above.  $\overline{(\tr\rho_A)^2}$ is again expressed as a sum over Ising configurations $\{s_\fp\}$ at polyhedra. It is straight-forward to compute the contribution at each $\{s_\fp\}$:
\be
\prod_{f\subset R_+\setminus \partial R_+}D_f^2\prod_{f\subset R_-\setminus \partial R_-}D_f^2\prod_{f\subset R_+\cap R_-} D_f\prod_{f\subset R_+\cap A}D^2_f\prod_{f\subset R_+\cap \bA}D_f^2\prod_{f\subset R_-\cap A}D_f\prod_{f\subset R_-\cap \bA}D_f
\ee
Now the dominant contribution comes from all $s_\fp=1$.
\be
\overline{(\tr\rho_A)^2}\simeq\prod_\fp\frac{1}{\dim(\ch_\fp)^2+\dim(\ch_\fp)}\prod_{f}D_f^2.
\ee 

The second Renyi entropy is obtained by the ratio
\be
e^{-S_2(A)}\simeq\frac{\overline{\tr_A\rho_A^2}}{\overline{(\tr\rho_A)^2}}=\sum_{\{\bar{s}_\fp\}}
\prod_{f\subset R_+\cap R_-} D_f^{-1}
\ee
It has been shown in Section \ref{bond} that $\ln D_f$ behaves as an area law in Eq.\Ref{arealaw}. Therefore
\be
e^{-S_2(A)}\simeq\sum_{\{\bar{s}_\fp\}}\exp\lt[-\frac{\b_0}{8\pi\g\ell_P^2}\sum_{f\subset R_+\cap R_-}\Ar_{f}\rt].
\ee

\begin{figure}[h]
\begin{center}
\includegraphics[width=6cm]{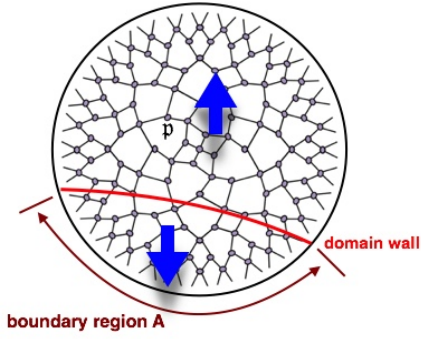}
\caption{An Ising configuration with a domain-wall separating two domains with opposite spins.}
\label{wall0}
\end{center}
\end{figure}

It is mentioned in Section \ref{bond} that $D_f$ of each face $f$ piecewisely is equivalent to the counting of states of a black hole in LQG \cite{GP2011}, where the black hole horizon area corresponds to $\Ar_{f}$. It is interesting to see that with spin-network at the quantum level, the problem of boundary entanglement entropy gets mapped to the problem of black hole entropy in the bulk. The domain-wall $R_+\cap R_-$ between $s_\fp=1$ and $s_\fp=-1$ (candidates of Ryu-Takayanagi surface) maps piecewisely to black hole horizons. It might be viewed as an quantum analog of the semiclassical derivation of Ryu-Takayanagi formula in e.g. \cite{LM2013,CHM2011}.

Consider zooming out to the macroscopic scale $L$ where each $\Ar_{f}$ is small, i.e. let $\ell_P^2\ll\Ar_f\ll L^2$ precisely the LQG semiclassical regime discussed at the beginning of Section \ref{EHM}. The continuum limit of the above formula is a path integral of Nambu-Goto action with surface tension $\mathrm{T}$:
\be
e^{-S_2(A)}\simeq\int [D \cs]\, e^{- \text{T} \Ar_\cs},\quad \text{T}=\frac{\b_0}{8\pi\g\ell_P^2}\label{NGaction}
\ee
where $\cs$ denotes an embedding of 2-surfaces such that $\partial\cs=\partial A$. The set of embeddings $\cs$ give the domain-walls $\equiv R_+\cap R_-$ of $\{\bar{s}_\fp\}$. The faces $f\subset R_+\cap R_-$ make a discretization of the surface $\cs$. $\sum_{\{\bar{s}_\fp\}}$ is a discrete version of summing over all embeddings, since different $\{\bar{s}_\fp\}$ given different discrete surfaces $R_+\cap R_-$. 

In the semiclassical limit $\text{T}=\frac{\b_0}{8\pi\g\ell_P^2}\to\infty$ \footnote{$\text{T}$ is a dimensionful quantity. So the proper way to understand the limit is that we zoom out to a larger scale such that $\ell_P\to 0$ can be taken. In other words, the Planck scale $\ell_P^2\ll \Ar_f$ the polyhedral lattice spacing. The Barbero-Immirzi parameter $\g$ is always assumed to be of order 1.}, the path-integral is dominated by the surfaces with locally minimal areas $\Ar_{\text{min}}$
\be
e^{-S_2(A)}\simeq\sum_{\text{minimal surfaces}}e^{-\frac{\b_0}{8\pi\g\ell_P^2} \Ar_{\text{min}}}. 
\ee
If we assume in the semi-classical geometry on $\Sig$, there is a unique global minimal surface area for 2-surfaces attached to $\partial A$ (e.g. $\text{AdS}_4$ geometry), then we obtain the Ryu-Takayanagi formula for second Renyi entropy
\be
S_2(A)\simeq\frac{\b_0}{8\pi\g\ell_P^2} \Ar_{\text{min}}.
\ee
Here the pre-factor $\frac{\b_0}{8\pi\g\ell_P^2}$ is identified to be IR value of $1/4 G_N$. 

\subsection{Higher Renyi Entropies}\label{higher}

In this subsection, we generalize the above second Renyi entropy computation to the random averaged higher Renyi entropies
\be
\overline{S_n(A)}\simeq\frac{1}{1-n}\ln\frac{\overline{\tr\rho_A^n}}{\overline{(\tr\rho_A)^n}}.
\ee
As an analog of Eq.\Ref{rho2}. $\tr\rho_A^n$ can be written as \footnote{It is straight-forward to check 
\be
&&\langle \mu^{(1)}_f|_A\langle \mu_f^{(1)}|_\bA\otimes\cdots\otimes\langle \mu^{(n)}_f|_A\langle \mu^{(n)}_f|_\bA\, \Big[(\rho\otimes\cdots\otimes\rho)\cc^{(n)}_A\Big]\,|\mu_f^{(1)}\rangle_A|\mu_f^{(1)}\rangle_\bA\otimes\cdots\otimes|\mu^{(n)}_f\rangle_A|\mu_f^{(n)}\rangle_\bA\nonumber\\
&=&\langle \mu^{(1)}_f|_A\langle \mu_f^{(1)}|_\bA\otimes\cdots\otimes\langle \mu^{(n)}_f|_A\langle \mu^{(n)}_f|_\bA\,\Big[\rho\otimes\cdots\otimes\rho\Big]\,|\mu_f^{(2)}\rangle_A|\mu_f^{(1)}\rangle_\bA\otimes\cdots\otimes|\mu^{(n)}_f\rangle_A|\mu_f^{(n-1)}\rangle_\bA\otimes|\mu^{(1)}_f\rangle_A|\mu_f^{(n)}\rangle_\bA\nonumber\\
&=&\Big(\langle \mu_f^{(1)}|_A\langle \mu^{(1)}_f|_\bA\Big)\,\rho\,\Big(|\mu^{(2)}_f\rangle_A|\mu^{(1)}_f\rangle_\bA\Big)\ 
\lt(\langle \mu^{(2)}_f|_A\langle \mu^{(2)}_f|_\bA\rt)\,\rho\,\lt(|\mu_f^{(3)}\rangle_A|\mu^{(2)}_f\rangle_\bA\rt)\cdots 
\lt(\langle \mu^{(n)}_f|_A\langle \mu^{(n)}_f|_\bA\rt)\,\rho\,\lt(|\mu_f^{(1)}\rangle_A|\mu^{(n)}_f\rangle\rt)\nonumber\\
&=&\langle \mu_f^{(1)}|\rho_A|\mu^{(2)}_f\rangle_A\ 
\langle \mu^{(2)}_f|\rho_A|\mu_f^{(3)}\rangle_A\cdots\langle \mu^{(n)}_f|\rho_A|\mu_f^{(1)}\rangle_A
\ee}
\be
\tr\rho_A^n=\langle \mu_f^{(1)}|\rho_A|\mu^{(2)}_f\rangle_A\ 
\langle \mu^{(2)}_f|\rho_A|\mu_f^{(3)}\rangle_A\cdots\langle \mu^{(n)}_f|\rho_A|\mu_f^{(1)}\rangle_A
=\tr_{\ch_{\partial}\otimes\cdots\otimes\ch_\partial}\lt[(\rho\otimes\cdots\otimes\rho)\cc^{(n)}_A\rt],
\ee
where repeating labels again means the summation over the labels. $\cc^{(n)}_A$ cyclicly permutes the states of region $A$, leaving the states of $\bA$ invariant:
\be
\cc^{(n)}_A\lt(|\mu_f^{(1)}\rangle_A|\mu_f^{(1)}\rangle_\bA\otimes\cdots\otimes|\mu^{(n)}_f\rangle_A|\mu_f^{(n)}\rangle_\bA\rt)=|\mu_f^{(2)}\rangle_A|\mu_f^{(1)}\rangle_\bA\otimes\cdots\otimes|\mu^{(n)}_f\rangle_A|\mu_f^{(n-1)}\rangle_\bA\otimes|\mu^{(1)}_f\rangle_A|\mu_f^{(n)}\rangle_\bA.
\ee
which reduces to $\cf_A$ at $n=2$.

Using the definition of $\rho$
\be
\tr\rho_A^n&=&\tr_{\ch_{\partial}\otimes\ch_\partial}\ \tr_{\ch_b\otimes\ch_E}^{\otimes n}\lt[\lt(\rho_P\otimes_\fp|\cv_\fp\rangle\langle \cv_\fp|\rt)^{\otimes n}\cc^{(n)}_A\rt].
\ee
To take the random average $\overline{\tr\rho_A^n}$ with random state $|\cv_\fp\rangle$, we use the following formula for the random average of n-fold tensor product, which generalizes Eq.\Ref{IF} \cite{Qi1,church}:
\be
\overline{\lt(|\cv_\fp\rangle\langle \cv_\fp|\rt)^{\otimes n}}=\frac{1}{C_{n,\fp}}\sum_{g_\fp\in\text{Sym}_n}g_{\fp}\in\ch_\fp^{\otimes n}\otimes\ch_\fp^*{}^{\otimes n}
\ee 
where the sum is over all permutations $g_\fp$ acting on $\ch_\fp^{\otimes n}$. The overall constant $C_{n,\fp}=\sum_{g_\fp\in\text{Sym}_n}\tr g_{\fp}=(\dim\ch_\fp+n-1)!/(\dim\ch_\fp-1)!$.

Inserting this result in $\tr\rho_A^n$, the average $\overline{\tr\rho_A^n}$ becomes a sum over all permutations $\{g_\fp\}$ at all polyhedra $\fp$, where each term associates to a choice of $g_\fp$ at each $\fp$:
\be
\langle \mu_f^{(1)}|_A\langle \mu_f^{(1)}|_\bA\langle\Phi_b|\langle E|\otimes\cdots\otimes\langle \mu^{(n)}_f|_A\langle \mu^{(n)}_f|_\bA\langle \Phi_b|\langle E|\prod_{\fp}g_\fp\ |\Phi_b\rangle|E\rangle|\mu^{(2)}_f\rangle_A|\mu^{(1)}_f\rangle_\bA\otimes\cdots\otimes |\Phi_b\rangle|E\rangle|\mu^{(1)}_f\rangle_A|\mu^{(n)}_f\rangle_\bA.
\ee
Firstly we compute the operator $\prod_{\fp}g_\fp$ acting on the right
\be
&&\prod_{\fp}g_\fp\ |\Phi_b\rangle|E\rangle|\mu^{(2)}_f\rangle_A|\mu^{(1)}_f\rangle_\bA\otimes\cdots\otimes |\Phi_b\rangle|E\rangle|\mu^{(1)}_f\rangle_A|\mu^{(n)}_f\rangle_\bA\nonumber\\
&=&\prod_{\fp}g_\fp\ \sum_{{\xi}^{(1)}_\fp,\mu^{(1)}_f}\Phi_{\vec{\xi}^{(1)},\mu^{(1)}_f,\{\mu_f^{(2)}\}_A,\{\mu_f^{(1)}\}_\bA}\otimes_\fp\lt|\xi_\fp^{\vec{\mu}}{}^{(1)},\mu^{(1)}_f,\{\mu_f^{(2)}\}_A,\{\mu_f^{(1)}\}_\bA\rag\otimes\nonumber\\
&&\cdots\otimes\sum_{{\xi}^{(n)}_\fp,\mu_f^{(n)}}\Phi_{\vec{\xi}^{(n)},{\mu}_f^{(n)},\{\mu_f^{(1)}\}_A,\{\mu^{(n)}_f\}_\bA}\otimes_\fp\lt|\xi_{\fp}^{\vec{\mu}'}{}^{(n)},\mu^{(n)}_f,\{\mu_f^{(1)}\}_A,\{\mu^{(n)}_f\}_\bA\rag\nonumber\\
&=&\sum_{{\xi}^{(1)}_\fp\cdots{\xi}^{(n)}_\fp,\mu^{(1)}_f\cdots\mu^{(n)}_f}
\Phi_{\vec{\xi}^{(1)},\mu^{(1)}_f,\{\mu_f^{(2)}\}_A,\{\mu_f^{(1)}\}_\bA}\cdots\Phi_{\vec{\xi}^{(n)},{\mu}_f^{(n)},\{\mu_f^{(1)}\}_A,\{\mu^{(n)}_f\}_\bA}\nonumber\\
&&\prod_{\fp}g_\fp\lt(\,\lt|\xi_\fp^{\vec{\mu}}{}^{(1)},\mu^{(1)}_f,\{\mu_f^{(2)}\}_A,\{\mu_f^{(1)}\}_\bA\rag\otimes\cdots\otimes\lt|\xi_{\fp}^{\vec{\mu}'}{}^{(n)},\mu^{(n)}_f,\{\mu_f^{(1)}\}_A,\{\mu^{(n)}_f\}_\bA\rag\,\rt)
\ee
$g_\fp$ only act on the states $| \xi_\fp^{\vec{\mu}}\rangle\otimes|\mu_f\rangle$ where the $\vec{\mu}$ labels of $\xi_\fp^{\vec{\mu}}$ coincide with the $\mu_f$ labels in $|\mu_f\rangle$. $g_\fp$ projects out the states which doesn't satisfy this coincidence. 

We take the inner product and compute the corresponding term in $\overline{\tr_A\rho_A^n}$ with a choice of $\{g_\fp\}$
\be
&&\sum_{{\zeta}^{(1)}_\fp\cdots{\zeta}^{(n)}_\fp,\nu^{(1)}_f\cdots\nu^{(n)}_f}\sum_{{\xi}^{(1)}_\fp\cdots{\xi}^{(n)}_\fp,\mu^{(1)}_f\cdots\mu^{(n)}_f}\sum_{\{\mu^{(1)}_f\cdots\mu^{(n)}_f\}_A}\sum_{\{\mu_f^{(1)}\cdots\mu^{(n)}_f\}_\bA}\nonumber\\
&&\Phi^*_{\vec{\zeta}^{(1)},\nu^{(1)}_f,\{\mu_f^{(1)}\}_A,\{\mu_f^{(1)}\}_\bA}\cdots\Phi^*_{\vec{\zeta}^{(n)},{\nu}_f^{(n)},\{\mu_f^{(n)}\}_A,\{\mu^{(n)}_f\}_\bA}
\Phi_{\vec{\xi}^{(1)},\mu^{(1)}_f,\{\mu_f^{(2)}\}_A,\{\mu_f^{(1)}\}_\bA}\cdots\Phi_{\vec{\xi}^{(n)},{\mu}_f^{(n)},\{\mu_f^{(1)}\}_A,\{\mu^{(n)}_f\}_\bA}\nonumber\\
&&\prod_{\fp}\delta_{\lt(\zeta^{(1)}_\fp\cdots\zeta_\fp^{(n)}\rt),\ g_\fp\lt(\xi^{(1)}_\fp\cdots\xi_\fp^{(n)}\rt)}\ 
\delta_{\lt(\nu^{(1)}_f\cdots\nu^{(n)}_f\rt),\ g_\fp\lt(\mu^{(1)}_f\cdots\mu^{(n)}_f\rt)}\ 
\delta_{\lt(\{\mu_f^{(1)}\}_A\cdots\{\mu_f^{(n)}\}_A\rt),\ g_\fp\lt(\{\mu^{(2)}_f\}_A\cdots\{\mu^{(n)}_f\}_A\{\mu^{(1)}_f\}_A\rt)}\ 
\delta_{\lt(\{\mu_f^{(1)}\}_\bA\cdots \{\mu_f^{(n)}\}_\bA\rt),\ g_{\fp}\lt(\{\mu^{(1)}_f\}_\bA\cdots\{\mu^{(n)}_f\}_\bA\rt)}.\label{termn}
\ee
We again focus on $\Phi_b$ satisfying Eqs.\Ref{factor} and \Ref{normalization}. Performing the sum $\sum_{{\zeta}^{(1)}_\fp\cdots{\zeta}^{(n)}_\fp}\sum_{{\xi}^{(1)}_\fp\cdots{\xi}^{(n)}_\fp}$ gives again an identity
\be
&&\sum_{{\zeta}^{(1)}_\fp\cdots{\zeta}^{(n)}_\fp}\sum_{{\xi}^{(1)}_\fp\cdots{\xi}^{(n)}_\fp}\prod_{\fp}\delta_{\lt(\zeta^{(1)}_\fp\cdots\zeta_\fp^{(n)}\rt),\ g_\fp\lt(\xi^{(1)}_\fp\cdots\xi_\fp^{(n)}\rt)}\nonumber\\
&&\Phi^*_{\vec{\zeta}^{(1)},\nu^{(1)}_f,\{\mu_f^{(1)}\}_A,\{\mu_f^{(1)}\}_\bA}\cdots\Phi^*_{\vec{\zeta}^{(n)},{\nu}_f^{(n)},\{\mu_f^{(n)}\}_A,\{\mu^{(n)}_f\}_\bA}
\Phi_{\vec{\xi}^{(1)},\mu^{(1)}_f,\{\mu_f^{(2)}\}_A,\{\mu_f^{(1)}\}_\bA}\cdots\Phi_{\vec{\xi}^{(n)},{\mu}_f^{(n)},\{\mu_f^{(1)}\}_A,\{\mu^{(n)}_f\}_\bA}=1.
\ee
Then Eq.\Ref{termn} simplifies to
\be
&&\sum_{\nu^{(1)}_f\cdots\nu^{(n)}_f}\sum_{\mu^{(1)}_f\cdots\mu^{(n)}_f}\sum_{\{\mu^{(1)}_f\cdots\mu^{(n)}_f\}_A}\sum_{\{\mu_f^{(1)}\cdots\mu^{(n)}_f\}_\bA}\nonumber\\
&&\prod_{\fp} \delta_{\lt(\nu^{(1)}_f\cdots\nu^{(n)}_f\rt),\ g_\fp\lt(\mu^{(1)}_f\cdots\mu^{(n)}_f\rt)}\ 
\delta_{\lt(\{\mu_f^{(1)}\}_A\cdots\{\mu_f^{(n)}\}_A\rt),\ g_\fp\lt(\{\mu^{(2)}_f\}_A\cdots\{\mu^{(n)}_f\}_A\{\mu^{(1)}_f\}_A\rt)}\ 
\delta_{\lt(\{\mu_f^{(1)}\}_\bA\cdots \{\mu_f^{(n)}\}_\bA\rt),\ g_{\fp}\lt(\{\mu^{(1)}_f\}_\bA\cdots\{\mu^{(n)}_f\}_\bA\rt)}, \label{termn1}
\ee
where the cut-offs $D_f$ have been imposed for the sums over $\mu_f^{(i)},\nu_f^{(i)}$ by $\Phi_b$.

The sum over $\{g_\fp\}$ is dominated by the contribution from $\{\bar{g}_\fp\}$ satisfying the following boundary condition
\be
\bar{g}_{\fp}\lt(\{\mu^{(1)}_f\}_\bA\cdots\{\mu^{(n)}_f\}_\bA\rt)&=&\lt(\{\mu^{(1)}_f\}_\bA\cdots\{\mu^{(n)}_f\}_\bA\rt)\nonumber\\
\bar{g}_\fp\lt(\{\mu^{(2)}_f\}_A\cdots\{\mu^{(n)}_f\}_A\{\mu^{(1)}_f\}_A\rt)&=&\lt(\{\mu_f^{(1)}\}_A\cdots\{\mu_f^{(n)}\}_A\rt)\label{b.c.}
\ee
i.e. for polyhedra $\fp$ connecting to $\partial\Sig$, $\bar{g}_\fp=I$ if $\fp$ is adjacent to $\bA$, while $\bar{g}_\fp=(\cc^{(n)})^{-1}$ if $\fp$ is adjacent to $A$. At each $\{\bar{g}_\fp\}$, Eq.\Ref{termn1} simplifies to
\be
&&\prod_{f\subset\partial\Sig}D^n_f\sum_{\nu^{(1)}_f\cdots\nu^{(n)}_f}\sum_{\mu^{(1)}_f\cdots\mu^{(n)}_f}\prod_{\fp} \delta_{\lt(\nu^{(1)}_f\cdots\nu^{(n)}_f\rt),\ \bar{g}_\fp\lt(\mu^{(1)}_f\cdots\mu^{(n)}_f\rt)}=\prod_{f\subset\partial\Sig}D^n_f\prod_{f\subset R_{\bar{g}}}D_f^n\prod_{\cs_{g,g'}}\prod_{f\subset \cs_{g,g'}}D_f^{\chi(\bar{g}^{-1}\bar{g}')}.
\ee
where we have denoted by $R_{g}$ the closed region of $\fp$'s with constant $g_\fp=g$. $R_g\cap R_{g'}\equiv\cs_{g,g'}$ denotes the interface (domain-wall) shared by $R_g, R_{g'}$ with two different permutations $g,g'$. $\chi(g)\leq n$ denotes the number of cycles in $g$, including the cycles of length one. $\chi(g)= n$ only when $g=I$.

As a result, we obtain
\be
\overline{\tr\rho_A^n}\simeq\prod_\fp\frac{1}{C_{n,\fp}}\prod_{f\subset\partial\Sig}D^n_f\sum_{\{\bar{g}_\fp\}}\prod_{f\subset R_{\bar{g}}}D_f^n\prod_{\cs_{g,g'}}\prod_{f\subset \cs_{g,g'}}D_f^{\chi(\bar{g}^{-1}\bar{g}')}.
\ee

Similarly, the dominant contribution of $\overline{(\tr_A\rho_A)^n}$ can be computed
\be
\overline{(\tr_A\rho_A)^n}\simeq \prod_\fp\frac{1}{C_{n,\fp}}\prod_{f}D^n_f
\ee
which corresponds to a constant $g_\fp=I$ everywhere in $\Sig$.

The $n$-th Renyi entropy is then given by
\be
e^{(1-n)S_n(A)}\simeq\sum_{\{\bar{g}_\fp\}}\prod_{\cs_{g,g'}}\prod_{f\subset \cs_{g,g'}}D_f^{\chi(\bar{g}^{-1}\bar{g}')-n}\simeq\sum_{\{\bar{g}_\fp\}}\exp\lt(\sum_{\cs_{\bar{g},\bar{g}'}}\lt[\chi(\bar{g}^{-1}\bar{g}')-n\rt]\frac{\b_0}{8\pi\g\ell_P^2}\,\Ar_{\cs_{\bar{g},\bar{g}'}}\rt).\label{expchi}
\ee
It corresponds to the partition function of a $\text{Sym}_n$-spin model similar to the one in \cite{Qi1}. The difference is that the non-uniform nearest-neighbor couplings are now interpreted as face areas $\ln D_f\simeq \frac{\b_0}{8\pi\g\ell_P^2}\,\Ar_{f}$.

The sum over configurations $\sum_{\{\bar{g}_\fp\}}$ can be understood as a sum over embedding of (discrete) surfaces $\cs_{\bar{g},\bar{g}'}$. By the same reasoning as the second Renyi entropy, the dominant contribution to $e^{(1-n)S_n(A)}$ comes from the configurations which minimize all $\Ar_{\cs_{\bar{g},\bar{g}'}}$. When the minimal surface attached to $\partial A$ is unique in the semiclassical geometry of $\Sig$, the sum is dominated by $\{\bar{g}_\fp\}$ which contains a single domain-wall $\cs_{\bar{g},\bar{g}'}$, separating $\bar{g}=I$ and $\bar{g}'=(\cc^{(n)})^{-1}$ consistent with the boundary condition Eq.\Ref{b.c.} (a proof of this statement is given in Appendix C of \cite{Qi1}, and is also provided here in Appendix \ref{DW} for completeness).

In the LQG semiclassical regime $\ell_P^2\ll\Ar_f\ll L^2$, we again can understand Eq.\Ref{expchi} (with single domain-wall) as a path integral of Nambu-Goto action as Eq.\Ref{NGaction}, with surface tension equals $(n-1)\frac{\b_0}{8\pi\g\ell_P^2}$. The single domain-wall of the minimial area corresponds to a configuration $\{\bar{g}_\fp\}_{\text{cri}}$, being the critical point of the path integral. At $\{\bar{g}_\fp\}_{\text{cri}}$, the Nambu-Goto action $\Ar_{\cs}=\Ar_{\text{min}}$ approaches its global minimum. This single domain wall is precisely the Ryu-Takayanagi surface with minimal surface area. As a result, in the limit $\ell_P\to 0$
\be
e^{(1-n)S_n(A)}\simeq e^{(1-n)\frac{\b_0}{8\pi\g\ell_P^2}\,\Ar_{\text{min}}},\quad S_n(A)\simeq \frac{\b_0}{8\pi\g\ell_P^2}\,\Ar_{\text{min}}\label{leading}
\ee
The resulting $S_n(A)$ is independent of $n$. 

The Von Neumann entropy $S(A)$ of reduced density matrix $\rho_A$ is given by $\lim_{n\to1}S_n(A)$. Since the leading contribution of $S_n(A)$ is independent of $n$, we have
\be
S(A)\simeq\frac{\b_0}{8\pi\g\ell_P^2}\,\Ar_{\text{min}},
\ee
which reproduces the Ryu-Takayanagi formula for entanglement entropy of boundary CFT. $\frac{\b_0}{8\pi\g\ell_P^2}$ is identified to be the IR value of $1/4 G_N$ in the bulk. It is consistent with what has been suggested in \cite{QGandBH,Ghosh:2012wq} from LQG perspective. From AdS/CFT perspective, $\frac{\b_0}{8\pi\g\ell_P^2}$ relates to the degree of freedom of CFT on the boundary. For instant, in $\text{AdS}_3/\text{CFT}_2$ it relates to the central charge $c=\frac{3L_{\text{AdS}}}{2G_N}$ \cite{brown1986}. In $\text{AdS}_4/\text{CFT}_3$ it relates to the free energy of CFT on $S^3$ \cite{Myers:2010tj,Bhattacharyya:2012tc}.

We have reproduced the Ryu-Takayanagi formula for the Renyi entropy $S_n(A)$ of arbitrary order $n$, and show $S_n(A)$ is independent of $n$ thus give the Von Neumann entropy $S(A)$ as above. This situation is the same as the computation with tensor networks in e.g. \cite{Qi1,Pastawski:2015qua}. However it is known that the Renyi entropy $S_n(A)$ of a CFT ground state has certain dependence on $n$ \cite{Calabrese:2004eu}. This mismatch is known in the literature for the tensor network state. Here we have the same issue because we relate LQG to tensor network in this work.



\section*{Acknowledgements}

The authors acknowledge Xiaoliang Qi, Michael Walter, and Zhao Yang for clarification about domain walls in $\text{Sym}_n$ model, and acknowledge Yidun Wan, Bei Zeng, and Jie Zhou for various stimulating discussions. MH acknowledges Wei Song and Yau Mathematical Sciences Center at Tsinghua University in Beijing, for the hospitality during his visits. MH also acknowledges support from the US National Science Foundation through grant PHY-1602867, and the Start-up Grant at Florida Atlantic University, USA. LYH acknowledges support from the Thousand Young Talents Program, and Fudan University in China.

\appendix

\section{Domain Walls in $\text{Sym}_n$ Spin Model}\label{DW}

\begin{figure}[h]
\begin{center}
\includegraphics[width=15cm]{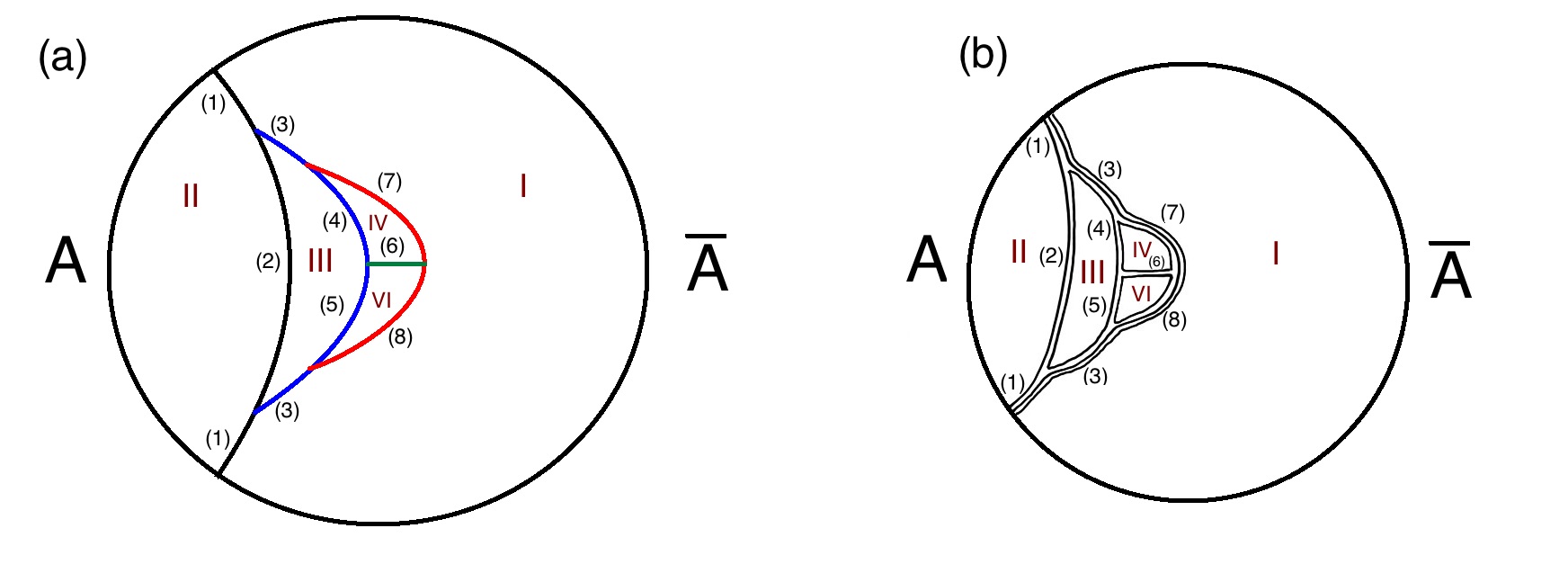}
\caption{(a) shows the space $\Sig$ with boundary $\partial\Sig$ divided into regions $A$ and $\bA$. $\Sig$ contains the domain-walls $(1),(2),\cdots, (8)$, which divide the bulk of $\Sig$ into regions I, II, $\cdots$, VI. The domain-wall $(1)\cup(2)$ is the unique surface with minimal area. Each bulk region associates a permutation $g_{I, II, \cdots, VI}$, with $g_I=I$ and $g_{II}=(\cc^{(n)})^{-1}$. Each domain-wall associates a number of cycles $\chi(g^{-1}g')=\chi(g'^{-1}g)$, with $g,g'$ on two sides of the domain-wall. (b) presents the domain-walls by using flow lines, because each domain-wall carries the Cayley weight $n-\chi(g^{-1}g')$ of a permutation $g^{-1}g'$. Since the flow lines only present the number of Cayley weight, there is no need to include any crossing of flow lines. The flow line picture is always planar.}
\label{walls}
\end{center}
\end{figure}

We come back to Eq.\Ref{leading}, and prove that the configuration with a single domain-wall indeed gives the leading contribution to $\sum_{\{g_\fp\}}$. Let's consider a more generic case shown in FIG.\ref{walls}(a), where more than one domain-walls are created in the bulk of $\Sig$. We are going to show that this configuration always contribute less than Eq.\Ref{leading} from a single domain-wall.

\begin{figure}[h]
\begin{center}
\includegraphics[width=6cm]{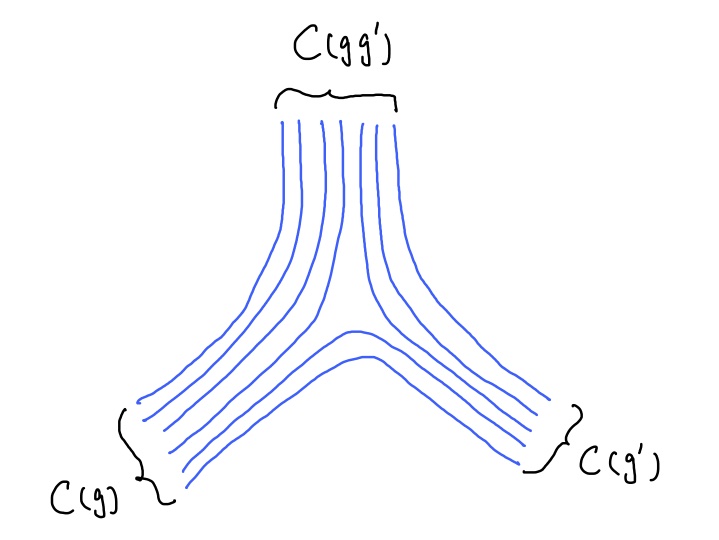}
\caption{A graphical presentation of the triangle inequality Eq.\Ref{triineq}.}
\label{flow}
\end{center}
\end{figure}

Given the multi-domain-wall configuration, each domain-wall carries the contribution proportional to
\be
-\lt[n-\chi(\bar{g}^{-1}\bar{g}')\rt]\Ar_{\cs_{\bar{g},\bar{g}'}}\label{chiA}
\ee
in Eq.\Ref{expchi}. For any permutation $g$, the number $n-\chi(g)\equiv C(g)$ is the Cayley weight of a permutation, which is defined by the minimum number of transpositions to achieve the permutation. It satisfies the triangle inequality \cite{Deza98metricson}
\be
n-\chi(gg')\leq \lt[n-\chi(g)\rt]+\lt[n-\chi(g')\rt].\label{triineq}
\ee
This inequality can be present graphically by using (planar) flow lines Fig.\ref{flow}.

There is a triangle inequality associated to each trivalent intersection of domain-walls\footnote{For higher valent intersection of domain-walls, the triangle inequality Eq.\Ref{triineq} guarantee that the flow line picture can still be used in the analysis. A higher valent intersection may be viewed dual to an abstract polygon. Each of the polygon edges is transverse to a unique domain-wall. The polygon can be made by gluing triangles with edge-lengths $n-\chi(g)$ (guaranteed by the triangle inequality Eq.\Ref{triineq}). Then we replace each triangle by the flow line picture FIG.\ref{flow}.}. For instance, in Fig.\ref{walls} at the intersection of (4), (5), and (6), we have
\be
n-\chi(g_{VI}^{-1}g_{IV})\leq \lt[n-\chi(g_{VI}^{-1} g_{III})\rt]+\lt[n-\chi(g_{III}^{-1}g_{IV})\rt]
\ee
This motivates us to present the domain-walls by using flow lines as in FIG.\ref{walls}(b) \footnote{The flow line are actually 2-surfaces when space $\Sig$ is 3-dimensional.}. Each segment of domain-wall $\cs_{g,g'}$ has the number $n-\chi(g^{-1}g')$ of flow lines, where each flow line carries the contribution $-\Ar_{\cs_{g,g'}}$ in Eqs.\Ref{expchi} and \Ref{chiA}. Then from FIG.\ref{walls}(b) it is not hard to see that the contribution of the multi-domain-wall configuration is less or equal to the single domain-wall configuration
\be
-\sum_{\cs_{\bar{g},\bar{g}'}}\lt[n-\chi(\bar{g}^{-1}\bar{g}')\rt]\Ar_{\cs_{\bar{g},\bar{g}'}}\leq -[n-1]\Ar_{\cs}\leq -[n-1]\Ar_{\text{min}}
\ee 
where $\cs$ is the domain-wall surface separating $I$ and $(\cc^{(n)})^{-1}$ satisfying $\partial\cs=\partial A$. In the continuum limit, viewing Eq.\Ref{expchi} as a path integral, the Ryu-Takayanagi surface is the critical point of the path integral where the Nambu-Goto action $\Ar_{\cs}=\Ar_{\text{min}}$ approaches its global minimum (We assume the minimal surface is unique in $\Sig$). Then Eq.\Ref{leading} is obtained in the semiclassical limit $\ell_P\to 0$.


\providecommand{\href}[2]{#2}\begingroup\raggedright\endgroup




\begin{thebibliography}{100}

\bibitem{Qi2}
X.-L. Qi, {\it {Exact holographic mapping and emergent space-time geometry}},
  \href{http://arxiv.org/abs/1309.6282}{{\tt arXiv:1309.6282}}.

\bibitem{Ryu:2006bv}
S.~Ryu and T.~Takayanagi, {\it {Holographic derivation of entanglement entropy
  from AdS/CFT}},  {\em Phys. Rev. Lett.} {\bf 96} (2006) 181602,
  [\href{http://arxiv.org/abs/hep-th/0603001}{{\tt hep-th/0603001}}].

\bibitem{Faulkner:2013yia}
T.~Faulkner, {\it {The Entanglement Renyi Entropies of Disjoint Intervals in
  AdS/CFT}},  \href{http://arxiv.org/abs/1303.7221}{{\tt arXiv:1303.7221}}.

\bibitem{Lewkowycz:2013nqa}
A.~Lewkowycz and J.~Maldacena, {\it {Generalized gravitational entropy}},  {\em
  JHEP} {\bf 08} (2013) 090, [\href{http://arxiv.org/abs/1304.4926}{{\tt
  arXiv:1304.4926}}].

\bibitem{Casini:2011kv}
H.~Casini, M.~Huerta, and R.~C. Myers, {\it {Towards a derivation of
  holographic entanglement entropy}},  {\em JHEP} {\bf 05} (2011) 036,
  [\href{http://arxiv.org/abs/1102.0440}{{\tt arXiv:1102.0440}}].

\bibitem{Sorkin:2014kta}
R.~D. Sorkin, {\it {1983 paper on entanglement entropy: "On the Entropy of the
  Vacuum outside a Horizon"}},  \href{http://arxiv.org/abs/1402.3589}{{\tt
  arXiv:1402.3589}}.

\bibitem{Callan:1994py}
C.~G. Callan, Jr. and F.~Wilczek, {\it {On geometric entropy}},  {\em Phys.
  Lett.} {\bf B333} (1994) 55--61,
  [\href{http://arxiv.org/abs/hep-th/9401072}{{\tt hep-th/9401072}}].

\bibitem{Faulkner:2013ana}
T.~Faulkner, A.~Lewkowycz, and J.~Maldacena, {\it {Quantum corrections to
  holographic entanglement entropy}},  {\em JHEP} {\bf 11} (2013) 074,
  [\href{http://arxiv.org/abs/1307.2892}{{\tt arXiv:1307.2892}}].

\bibitem{Jafferis:2015del}
D.~L. Jafferis, A.~Lewkowycz, J.~Maldacena, and S.~J. Suh, {\it {Relative
  entropy equals bulk relative entropy}},  {\em JHEP} {\bf 06} (2016) 004,
  [\href{http://arxiv.org/abs/1512.06431}{{\tt arXiv:1512.06431}}].

\bibitem{Almheiri:2014lwa}
A.~Almheiri, X.~Dong, and D.~Harlow, {\it {Bulk Locality and Quantum Error
  Correction in AdS/CFT}},  {\em JHEP} {\bf 04} (2015) 163,
  [\href{http://arxiv.org/abs/1411.7041}{{\tt arXiv:1411.7041}}].

\bibitem{Harlow:2016vwg}
D.~Harlow, {\it {The Ryu-Takayanagi Formula from Quantum Error Correction}},
  \href{http://arxiv.org/abs/1607.03901}{{\tt arXiv:1607.03901}}.

\bibitem{Bianchi:2012ev}
E.~Bianchi and R.~C. Myers, {\it {On the Architecture of Spacetime Geometry}},
  {\em Class. Quant. Grav.} {\bf 31} (2014) 214002,
  [\href{http://arxiv.org/abs/1212.5183}{{\tt arXiv:1212.5183}}].

\bibitem{Swingle:2009bg}
B.~Swingle, {\it {Entanglement Renormalization and Holography}},  {\em Phys.
  Rev.} {\bf D86} (2012) 065007, [\href{http://arxiv.org/abs/0905.1317}{{\tt
  arXiv:0905.1317}}].

\bibitem{Orus:2014poa}
R.~Orus, {\it {Advances on Tensor Network Theory: Symmetries, Fermions,
  Entanglement, and Holography}},  {\em Eur. Phys. J.} {\bf B87} (2014) 280,
  [\href{http://arxiv.org/abs/1407.6552}{{\tt arXiv:1407.6552}}].

\bibitem{Bridgeman:2016dhh}
J.~C. Bridgeman and C.~T. Chubb, {\it {Hand-waving and Interpretive Dance: An
  Introductory Course on Tensor Networks}},
  \href{http://arxiv.org/abs/1603.03039}{{\tt arXiv:1603.03039}}.

\bibitem{2008PhRvL.101k0501V}
G.~{Vidal}, {\it {Class of Quantum Many-Body States That Can Be Efficiently
  Simulated}},  {\em Physical Review Letters} {\bf 101} (Sept., 2008) 110501,
  [\href{http://arxiv.org/abs/quant-ph/0610099}{{\tt quant-ph/0610099}}].

\bibitem{2013PhRvB..88k5147S}
S.~Singh and G.~Vidal, {\it Global symmetries in tensor network states:
  Symmetric tensors versus minimal bond dimension},  {\em Phys. Rev. B} {\bf
  88} (Sep, 2013) 115147.

\bibitem{Pastawski:2015qua}
F.~Pastawski, B.~Yoshida, D.~Harlow, and J.~Preskill, {\it {Holographic quantum
  error-correcting codes: Toy models for the bulk/boundary correspondence}},
  {\em JHEP} {\bf 06} (2015) 149, [\href{http://arxiv.org/abs/1503.06237}{{\tt
  arXiv:1503.06237}}].

\bibitem{Qi1}
P.~Hayden, S.~Nezami, X.-L. Qi, N.~Thomas, M.~Walter, and Z.~Yang, {\it
  {Holographic duality from random tensor networks}},
  \href{http://arxiv.org/abs/1601.01694}{{\tt arXiv:1601.01694}}.

\bibitem{book}
T.~Thiemann, {\em Modern Canonical Quantum General Relativity}.
\newblock Cambridge University Press, 2007.

\bibitem{rovelli2014covariant}
C.~Rovelli and F.~Vidotto, {\em Covariant Loop Quantum Gravity: An Elementary
  Introduction to Quantum Gravity and Spinfoam Theory}.
\newblock Cambridge Monographs on Mathematical Physics. Cambridge University
  Press, 2014.

\bibitem{review1}
A.~Ashtekar and J.~Lewandowski, {\it {Background independent quantum gravity: A
  Status report}},  {\em Class.Quant.Grav.} {\bf 21} (2004) R53,
  [\href{http://arxiv.org/abs/gr-qc/0404018}{{\tt gr-qc/0404018}}].

\bibitem{review}
M.~Han, W.~Huang, and Y.~Ma, {\it {Fundamental structure of loop quantum
  gravity}},  {\em Int.J.Mod.Phys.} {\bf D16} (2007) 1397--1474,
  [\href{http://arxiv.org/abs/gr-qc/0509064}{{\tt gr-qc/0509064}}].

\bibitem{Bodendorfer:2011nx}
N.~Bodendorfer, T.~Thiemann, and A.~Thurn, {\it {New Variables for Classical
  and Quantum Gravity in all Dimensions III. Quantum Theory}},  {\em Class.
  Quant. Grav.} {\bf 30} (2013) 045003,
  [\href{http://arxiv.org/abs/1105.3705}{{\tt arXiv:1105.3705}}].

\bibitem{Ashtekar:1986yd}
A.~Ashtekar, {\it {New Variables for Classical and Quantum Gravity}},  {\em
  Phys. Rev. Lett.} {\bf 57} (1986) 2244--2247.

\bibitem{Rovelli1988}
C.~Rovelli and L.~Smolin, {\it {Knot Theory and Quantum Gravity}},  {\em
  Physical Review Letters} {\bf 61} (Sept., 1988) 1155--1158.

\bibitem{Ashtekar:1991kc}
A.~Ashtekar and C.~J. Isham, {\it {Representations of the holonomy algebras of
  gravity and nonAbelian gauge theories}},  {\em Class. Quant. Grav.} {\bf 9}
  (1992) 1433--1468, [\href{http://arxiv.org/abs/hep-th/9202053}{{\tt
  hep-th/9202053}}].

\bibitem{uniqueness}
J.~Lewandowski, A.~Okolow, H.~Sahlmann, and T.~Thiemann, {\it {Uniqueness of
  diffeomorphism invariant states on holonomy-flux algebras}},  {\em Commun.
  Math. Phys.} {\bf 267} (2006) 703--733,
  [\href{http://arxiv.org/abs/gr-qc/0504147}{{\tt gr-qc/0504147}}].

\bibitem{Fleischhack:2007mj}
C.~Fleischhack, {\it {Kinematical uniqueness of loop quantum gravity}}, \href{http://arxiv.org/abs/1505.04404}{{\tt arXiv:1505.04404}}.

\bibitem{Rovelli1995}
C.~Rovelli and L.~Smolin, {\it {Discreteness of area and volume in quantum
  gravity}},  {\em Nuclear Physics B} {\bf 442} (May, 1995) 593--619.

\bibitem{ALarea}
A.~Ashtekar and J.~Lewandowski, {\it {Quantum theory of geometry. 1: Area
  operators}},  {\em Class.Quant.Grav.} {\bf 14} (1997) A55--A82,
  [\href{http://arxiv.org/abs/gr-qc/9602046}{{\tt gr-qc/9602046}}].

\bibitem{ALvolume}
A.~Ashtekar and J.~Lewandowski, {\it {Quantum theory of geometry. 2. Volume
  operators}},  {\em Adv.Theor.Math.Phys.} {\bf 1} (1998) 388--429,
  [\href{http://arxiv.org/abs/gr-qc/9711031}{{\tt gr-qc/9711031}}].

\bibitem{Thiemann:1996at}
T.~Thiemann, {\it {A Length operator for canonical quantum gravity}},  {\em J.
  Math. Phys.} {\bf 39} (1998) 3372--3392,
  [\href{http://arxiv.org/abs/gr-qc/9606092}{{\tt gr-qc/9606092}}].

\bibitem{Bianchi:2008es}
E.~Bianchi, {\it {The Length operator in Loop Quantum Gravity}},  {\em Nucl.
  Phys.} {\bf B807} (2009) 591--624,
  [\href{http://arxiv.org/abs/0806.4710}{{\tt arXiv:0806.4710}}].

\bibitem{Ma:2010fy}
Y.~Ma, C.~Soo, and J.~Yang, {\it {New length operator for loop quantum
  gravity}},  {\em Phys. Rev.} {\bf D81} (2010) 124026,
  [\href{http://arxiv.org/abs/1004.1063}{{\tt arXiv:1004.1063}}].

\bibitem{Ma:2000au}
Y.~Ma and Y.~Ling, {\it {The Q-hat operator for canonical quantum gravity}},
  {\em Phys. Rev.} {\bf D62} (2000) 104021,
  [\href{http://arxiv.org/abs/gr-qc/0005117}{{\tt gr-qc/0005117}}].

\bibitem{Ghosh:2013iwa}
A.~Ghosh, K.~Noui, and A.~Perez, {\it {Statistics, holography, and black hole
  entropy in loop quantum gravity}},  {\em Phys. Rev.} {\bf D89} (2014), no.~8
  084069, [\href{http://arxiv.org/abs/1309.4563}{{\tt arXiv:1309.4563}}].

\bibitem{hanBH}
M.~Han, {\it {Black hole entropy in loop quantum gravity, analytic
  continuation, and dual holography}},
  \href{http://arxiv.org/abs/1402.2084}{{\tt arXiv:1402.2084}}.

\bibitem{Han:2016fgh}
M.~Han and M.~Zhang, {\it {On Spinfoams Near a Classical Curvature
  Singularity}},  \href{http://arxiv.org/abs/1606.02826}{{\tt
  arXiv:1606.02826}}.

\bibitem{lowE}
M.~Han, {\it {Covariant loop quantum gravity, low energy perturbation theory,
  and Einstein gravity with high curvature UV corrections}},  {\em Phys.Rev.}
  {\bf D89} (2014) 124001, [\href{http://arxiv.org/abs/1308.4063}{{\tt
  arXiv:1308.4063}}].

\bibitem{Sahlmann:2001nv}
H.~Sahlmann, T.~Thiemann, and O.~Winkler, {\it {Coherent states for canonical
  quantum general relativity and the infinite tensor product extension}},  {\em
  Nucl. Phys.} {\bf B606} (2001) 401--440,
  [\href{http://arxiv.org/abs/gr-qc/0102038}{{\tt gr-qc/0102038}}].

\bibitem{Brunnemann:2007ca}
J.~Brunnemann and D.~Rideout, {\it {Properties of the volume operator in loop
  quantum gravity. I. Results}},  {\em Class. Quant. Grav.} {\bf 25} (2008)
  065001, [\href{http://arxiv.org/abs/0706.0469}{{\tt arXiv:0706.0469}}].

\bibitem{Bianchi:2011ub}
E.~Bianchi and H.~M. Haggard, {\it {Discreteness of the volume of space from
  Bohr-Sommerfeld quantization}},  {\em Phys. Rev. Lett.} {\bf 107} (2011)
  011301, [\href{http://arxiv.org/abs/1102.5439}{{\tt arXiv:1102.5439}}].

\bibitem{GP2011}
A.~Ghosh and A.~Perez, {\it {Black hole entropy and isolated horizons
  thermodynamics}},  {\em Phys. Rev. Lett.} {\bf 107} (2011) 241301,
  [\href{http://arxiv.org/abs/1107.1320}{{\tt arXiv:1107.1320}}]. [Erratum:
  Phys. Rev. Lett.108,169901(2012)].

\bibitem{Ghosh:2004wq}
A.~Ghosh and P.~Mitra, {\it {An Improved lower bound on black hole entropy in
  the quantum geometry approach}},  {\em Phys. Lett.} {\bf B616} (2005)
  114--117, [\href{http://arxiv.org/abs/gr-qc/0411035}{{\tt gr-qc/0411035}}].

\bibitem{QGandBH}
J.~F. Barbero~G. and A.~Perez, {\it {Quantum Geometry and Black Holes}},
  \href{http://arxiv.org/abs/1501.02963}{{\tt arXiv:1501.02963}}.

\bibitem{Ghosh:2012wq}
A.~Ghosh and A.~Perez, {\it {The scaling of black hole entropy in loop quantum
  gravity}},  \href{http://arxiv.org/abs/1210.2252}{{\tt arXiv:1210.2252}}.

\bibitem{randomrev}
B.~Collins and I.~Nechita, {\it Random matrix techniques in quantum information
  theory},  {\em Journal of Mathematical Physics} {\bf 57} (2016), no.~1.

\bibitem{hayden}
P.~Hayden, {\it Entanglement in random subspaces},  {\em AIP Conference
  Proceedings} {\bf 734} (2004), no.~1.

\bibitem{Page:1993df}
D.~N. Page, {\it {Average entropy of a subsystem}},  {\em Phys. Rev. Lett.}
  {\bf 71} (1993) 1291--1294, [\href{http://arxiv.org/abs/gr-qc/9305007}{{\tt
  gr-qc/9305007}}].

\bibitem{Bodendorfer:2013sja}
N.~Bodendorfer, {\it {Black hole entropy from loop quantum gravity in higher
  dimensions}},  {\em Phys. Lett.} {\bf B726} (2013) 887--891,
  [\href{http://arxiv.org/abs/1307.5029}{{\tt arXiv:1307.5029}}].

\bibitem{future}
M.~Han and L.-Y. Hung \href{http://arxiv.org/abs/In preparation}{{\tt In
  preparation}}.

\bibitem{Markopoulou:1999iq}
F.~Markopoulou and L.~Smolin, {\it {Holography in a quantum space-time}},
  \href{http://arxiv.org/abs/hep-th/9910146}{{\tt hep-th/9910146}}.

\bibitem{Ling:2000ss}
Y.~Ling and L.~Smolin, {\it {Holographic formulation of quantum supergravity}},
   {\em Phys. Rev.} {\bf D63} (2001) 064010,
  [\href{http://arxiv.org/abs/hep-th/0009018}{{\tt hep-th/0009018}}].

\bibitem{Han:2014xna}
M.~Han, {\it {Black Hole Entropy in Loop Quantum Gravity, Analytic
  Continuation, and Dual Holography}},
  \href{http://arxiv.org/abs/1402.2084}{{\tt arXiv:1402.2084}}.

\bibitem{Ghosh:2014rra}
A.~Ghosh and D.~Pranzetti, {\it {CFT/Gravity Correspondence on the Isolated
  Horizon}},  {\em Nucl. Phys.} {\bf B889} (2014) 1--24,
  [\href{http://arxiv.org/abs/1405.7056}{{\tt arXiv:1405.7056}}].

\bibitem{Zuo:2016ezr}
F.~Zuo, {\it {A note on the architecture of spacetime geometry}},
  \href{http://arxiv.org/abs/1607.05866}{{\tt arXiv:1607.05866}}.

\bibitem{Bonzom:2015ans}
V.~Bonzom and B.~Dittrich, {\it {3D holography: from discretum to continuum}},
  {\em JHEP} {\bf 03} (2016) 208, [\href{http://arxiv.org/abs/1511.05441}{{\tt
  arXiv:1511.05441}}].

\bibitem{hanSUSY}
M.~Han, {\it {4d Quantum Geometry from 3d Supersymmetric Gauge Theory and
  Holomorphic Block}},  {\em JHEP} {\bf 01} (2016) 065,
  [\href{http://arxiv.org/abs/1509.00466}{{\tt arXiv:1509.00466}}].

\bibitem{Yang:2008th}
J.~Yang and Y.~Ma, {\it {Quasi-Local Energy in Loop Quantum Gravity}},  {\em
  Phys. Rev.} {\bf D80} (2009) 084027,
  [\href{http://arxiv.org/abs/0812.3554}{{\tt arXiv:0812.3554}}].

\bibitem{Freidel:2015gpa}
L.~Freidel and A.~Perez, {\it {Quantum gravity at the corner}},
  \href{http://arxiv.org/abs/1507.02573}{{\tt arXiv:1507.02573}}.

\bibitem{Smolin:2016edy}
L.~Smolin, {\it {Holographic relations in loop quantum gravity}},
  \href{http://arxiv.org/abs/1608.02932}{{\tt arXiv:1608.02932}}.

\bibitem{Perez:2014ura}
A.~Perez, {\it {Statistical and entanglement entropy for black holes in quantum
  geometry}},  {\em Phys. Rev.} {\bf D90} (2014), no.~8 084015,
  [\href{http://arxiv.org/abs/1405.7287}{{\tt arXiv:1405.7287}}]. [Addendum:
  Phys. Rev.D90,no.8,089907(2014)].

\bibitem{Bianchi:2015fra}
E.~Bianchi, L.~Hackl, and N.~Yokomizo, {\it {Entanglement entropy of squeezed
  vacua on a lattice}},  {\em Phys. Rev.} {\bf D92} (2015), no.~8 085045,
  [\href{http://arxiv.org/abs/1507.01567}{{\tt arXiv:1507.01567}}].

\bibitem{Bianchi:2014bma}
E.~Bianchi, T.~De~Lorenzo, and M.~Smerlak, {\it {Entanglement entropy
  production in gravitational collapse: covariant regularization and solvable
  models}},  {\em JHEP} {\bf 06} (2015) 180,
  [\href{http://arxiv.org/abs/1409.0144}{{\tt arXiv:1409.0144}}].

\bibitem{Bodendorfer:2014fua}
N.~Bodendorfer, {\it {A note on entanglement entropy and quantum geometry}},
  {\em Class. Quant. Grav.} {\bf 31} (2014), no.~21 214004,
  [\href{http://arxiv.org/abs/1402.1038}{{\tt arXiv:1402.1038}}].

\bibitem{Delcamp:2016eya}
C.~Delcamp, B.~Dittrich, and A.~Riello, {\it {On entanglement entropy in
  non-Abelian lattice gauge theory and 3D quantum gravity}},
  \href{http://arxiv.org/abs/1609.04806}{{\tt arXiv:1609.04806}}.

\bibitem{Livine:2005mw}
E.~R. Livine and D.~R. Terno, {\it {Quantum black holes: Entropy and
  entanglement on the horizon}},  {\em Nucl. Phys.} {\bf B741} (2006) 131--161,
  [\href{http://arxiv.org/abs/gr-qc/0508085}{{\tt gr-qc/0508085}}].

\bibitem{Donnelly:2016auv}
W.~Donnelly and L.~Freidel, {\it {Local subsystems in gauge theory and
  gravity}},  {\em JHEP} {\bf 09} (2016) 102,
  [\href{http://arxiv.org/abs/1601.04744}{{\tt arXiv:1601.04744}}].

\bibitem{Donnelly:2008vx}
W.~Donnelly, {\it {Entanglement entropy in loop quantum gravity}},  {\em Phys.
  Rev.} {\bf D77} (2008) 104006, [\href{http://arxiv.org/abs/0802.0880}{{\tt
  arXiv:0802.0880}}].

\bibitem{Donnelly:2011hn}
W.~Donnelly, {\it {Decomposition of entanglement entropy in lattice gauge
  theory}},  {\em Phys. Rev.} {\bf D85} (2012) 085004,
  [\href{http://arxiv.org/abs/1109.0036}{{\tt arXiv:1109.0036}}].

\bibitem{Dittrich:2014mxa}
B.~Dittrich, S.~Mizera, and S.~Steinhaus, {\it {Decorated tensor network
  renormalization for lattice gauge theories and spin foam models}},  {\em New
  J. Phys.} {\bf 18} (2016), no.~5 053009,
  [\href{http://arxiv.org/abs/1409.2407}{{\tt arXiv:1409.2407}}].

\bibitem{Dittrich:2016tys}
B.~Dittrich, E.~Schnetter, C.~J. Seth, and S.~Steinhaus, {\it {Coarse graining
  flow of spin foam intertwiners}},
  \href{http://arxiv.org/abs/1609.02429}{{\tt arXiv:1609.02429}}.

\bibitem{Anza:2016fix}
F.~Anz\`a and G.~Chirco, {\it {Quantum typicality in spin network states of
  quantum geometry}},  \href{http://arxiv.org/abs/1605.04946}{{\tt
  arXiv:1605.04946}}.

\bibitem{Dittrich:2013bza}
B.~Dittrich, M.~Martín-Benito, and E.~Schnetter, {\it {Coarse graining of spin
  net models: dynamics of intertwiners}},  {\em New J. Phys.} {\bf 15} (2013)
  103004, [\href{http://arxiv.org/abs/1306.2987}{{\tt arXiv:1306.2987}}].

\bibitem{Bodendorfer:2016tky}
N.~Bodendorfer, {\it {State refinements and coarse graining in a full theory
  embedding of loop quantum cosmology}},
  \href{http://arxiv.org/abs/1607.06227}{{\tt arXiv:1607.06227}}.

\bibitem{Livine:2013gna}
E.~R. Livine, {\it {Deformation Operators of Spin Networks and
  Coarse-Graining}},  {\em Class. Quant. Grav.} {\bf 31} (2014) 075004,
  [\href{http://arxiv.org/abs/1310.3362}{{\tt arXiv:1310.3362}}].

\bibitem{Bahr:2012qj}
B.~Bahr, B.~Dittrich, F.~Hellmann, and W.~Kaminski, {\it {Holonomy Spin Foam
  Models: Definition and Coarse Graining}},  {\em Phys. Rev.} {\bf D87} (2013),
  no.~4 044048, [\href{http://arxiv.org/abs/1208.3388}{{\tt arXiv:1208.3388}}].

\bibitem{Livine:2006xk}
E.~R. Livine and D.~R. Terno, {\it {Reconstructing quantum geometry from
  quantum information: Area renormalisation, coarse-graining and entanglement
  on spin networks}},  \href{http://arxiv.org/abs/gr-qc/0603008}{{\tt
  gr-qc/0603008}}.

\bibitem{holst}
S.~Holst, {\it {Barbero's Hamiltonian derived from a generalized
  Hilbert-Palatini action}},  {\em Phys.Rev.} {\bf D53} (1996) 5966--5969,
  [\href{http://arxiv.org/abs/gr-qc/9511026}{{\tt gr-qc/9511026}}].

\bibitem{barbero}
J.~F. Barbero~G., {\it {Real Ashtekar variables for Lorentzian signature space
  times}},  {\em Phys.Rev.} {\bf D51} (1995) 5507--5510,
  [\href{http://arxiv.org/abs/gr-qc/9410014}{{\tt gr-qc/9410014}}].

\bibitem{Marolf:1994cj}
D.~Marolf and J.~M. Mourao, {\it {On the support of the Ashtekar-Lewandowski
  measure}},  {\em Commun. Math. Phys.} {\bf 170} (1995) 583--606,
  [\href{http://arxiv.org/abs/hep-th/9403112}{{\tt hep-th/9403112}}].

\bibitem{Ashtekar:1993wf}
A.~Ashtekar and J.~Lewandowski, {\it {Representation theory of analytic
  holonomy C* algebras}},  \href{http://arxiv.org/abs/gr-qc/9311010}{{\tt
  gr-qc/9311010}}.

\bibitem{shape}
E.~Bianchi, P.~Dona, and S.~Speziale, {\it {Polyhedra in loop quantum
  gravity}},  {\em Phys.Rev.} {\bf D83} (2011) 044035,
  [\href{http://arxiv.org/abs/1009.3402}{{\tt arXiv:1009.3402}}].

\bibitem{CF}
F.~Conrady and L.~Freidel, {\it {Quantum geometry from phase space reduction}},
   {\em J.Math.Phys.} {\bf 50} (2009) 123510,
  [\href{http://arxiv.org/abs/0902.0351}{{\tt arXiv:0902.0351}}].

\bibitem{QSF}
M.~Han, {\it {4-dimensional spin-foam model with quantum Lorentz group}},  {\em
  J.Math.Phys.} {\bf 52} (2011) 072501,
  [\href{http://arxiv.org/abs/1012.4216}{{\tt arXiv:1012.4216}}].

\bibitem{QSF1}
W.~J. Fairbairn and C.~Meusburger, {\it {Quantum deformation of two
  four-dimensional spin foam models}},  {\em J.Math.Phys.} {\bf 53} (2012)
  022501, [\href{http://arxiv.org/abs/1012.4784}{{\tt arXiv:1012.4784}}].

\bibitem{3dblockHHKR}
H.~M. Haggard, M.~Han, W.~Kaminski, and A.~Riello, {\it {SL(2,C) Chern-Simons
  Theory, Flat Connections, and Four-dimensional Quantum Geometry}},
  \href{http://arxiv.org/abs/1512.07690}{{\tt arXiv:1512.07690}}.

\bibitem{HHKRshort}
H.~M. Haggard, M.~Han, W.~Kaminski, and A.~Riello, {\it {Four-dimensional
  Quantum Gravity with a Cosmological Constant from Three-dimensional
  Holomorphic Blocks}},  {\em Phys. Lett.} {\bf B752} (2016) 258--262,
  [\href{http://arxiv.org/abs/1509.00458}{{\tt arXiv:1509.00458}}].

\bibitem{HHKR}
H.~M. Haggard, M.~Han, W.~Kaminski, and A.~Riello, {\it {SL(2,C) Chern-Simons
  Theory, a non-Planar Graph Operator, and 4D Loop Quantum Gravity with a
  Cosmological Constant: Semiclassical Geometry}},  {\em Nucl. Phys.} {\bf
  B900} (2015) 1--79, [\href{http://arxiv.org/abs/1412.7546}{{\tt
  arXiv:1412.7546}}].

\bibitem{polymer}
E.~Bianchi, {\it {Black Hole Entropy, Loop Gravity, and Polymer Physics}},
  {\em Class. Quant. Grav.} {\bf 28} (2011) 114006,
  [\href{http://arxiv.org/abs/1011.5628}{{\tt arXiv:1011.5628}}].

\bibitem{pathria}
R.~Pathria and P.~Beale, {\em Statistical Mechanics}.
\newblock Elsevier Science, 1996.

\bibitem{church}
A.~W. Harrow, {\it {The church of the symmetric subspace}},
  \href{http://arxiv.org/abs/1308.6595}{{\tt arXiv:1308.6595}}.

\bibitem{LM2013}
A.~Lewkowycz and J.~Maldacena, {\it {Generalized gravitational entropy}},  {\em
  JHEP} {\bf 08} (2013) 090, [\href{http://arxiv.org/abs/1304.4926}{{\tt
  arXiv:1304.4926}}].

\bibitem{CHM2011}
H.~Casini, M.~Huerta, and R.~C. Myers, {\it {Towards a derivation of
  holographic entanglement entropy}},  {\em JHEP} {\bf 05} (2011) 036,
  [\href{http://arxiv.org/abs/1102.0440}{{\tt arXiv:1102.0440}}].

\bibitem{brown1986}
J.~D. Brown and M.~Henneaux, {\it Central charges in the canonical realization
  of asymptotic symmetries: an example from three-dimensional gravity},  {\em
  Comm. Math. Phys.} {\bf 104} (1986), no.~2 207--226.

\bibitem{Myers:2010tj}
R.~C. Myers and A.~Sinha, {\it {Holographic c-theorems in arbitrary
  dimensions}},  {\em JHEP} {\bf 01} (2011) 125,
  [\href{http://arxiv.org/abs/1011.5819}{{\tt arXiv:1011.5819}}].

\bibitem{Bhattacharyya:2012tc}
A.~Bhattacharyya, L.-Y. Hung, K.~Sen, and A.~Sinha, {\it {On c-theorems in
  arbitrary dimensions}},  {\em Phys. Rev.} {\bf D86} (2012) 106006,
  [\href{http://arxiv.org/abs/1207.2333}{{\tt arXiv:1207.2333}}].

\bibitem{Calabrese:2004eu}
P.~Calabrese and J.~L. Cardy, {\it {Entanglement entropy and quantum field
  theory}},  {\em J. Stat. Mech.} {\bf 0406} (2004) P06002,
  [\href{http://arxiv.org/abs/hep-th/0405152}{{\tt hep-th/0405152}}].

\bibitem{Deza98metricson}
M.~Deza and T.~Huang, {\it Metrics on permutations, a survey},  {\em Journal of
  Combinatorics, Information and System Sciences} (1998).

\end{thebibliography}

\end{document}